\documentclass[12pt,letterpaper]{report}          
\usepackage[intlimits]{amsmath}
\usepackage{amsfonts,amssymb}
\DeclareSymbolFontAlphabet{\mathbb}{AMSb}
\usepackage{apalike}
\usepackage{float}
\usepackage[bf]{caption}       
\usepackage{fancyhdr}
\usepackage{fancybox}
\usepackage{ifthen}
\usepackage{bu_ece_thesis}
\usepackage{url}
\usepackage{lscape,afterpage}
\usepackage{xspace}
\usepackage{xcolor}
\usepackage{epstopdf} 
\usepackage{subfig}
\usepackage{tikz}

\usepackage{graphicx}
\usepackage{appendix}

\begin{document}


\title{Simplicial Lattice Study of the 2d Ising CFT}

\author{Evan Owen}

\degree=2

\prevdegrees{B.S. Mechanical Engineering, Cal Poly, San Luis Obispo, 2009\\
	M.S. Physics, San Francisco State University, 2019}

\department{Department of Physics}

\defenseyear{2023}
\degreeyear{2023}

\reader{First}{Richard C. Brower, Ph.D.}{Professor of Electrical and Computer Engineering}
\reader{Second}{Emmanuel Katz, Ph.D.}{Professor of Physics}
\reader{Third}{Ethan T. Neil, Ph.D.}{Associate Professor of Physics\\University of Colorado, Boulder}

\numadvisors=1
\majorprof{Richard C. Brower, Ph.D.}{{Professor of Electrical and Computer Engineering}}




\maketitle
\cleardoublepage

\copyrightpage
\cleardoublepage

\approvalpagewithcomment
\cleardoublepage


\newpage
\section*{\centerline{Acknowledgments}}

A lot of people supported me during grad school. My academic collaborators were obviously involved with all of the research work that I've done, but here I’d like to acknowledge some of the people outside of academia who made it possible for me to finish my Ph.D. Without the extraordinary support of my friends and family this work would not have been possible.

My dad, Jay, who built model rockets with me and taught me about transistors and capacitors. My mom, Carol, who taught me to stand up for myself and not to settle for less than I deserve. My high school physics teacher, Mr. Fottrell, who first introduced me to the connection between electricity and magnetism. My friends from the Master's program at SF State: Priti, Katie, Nikki, Audrey, Sarah, Wilder, and Stephen, who were the best colleagues anyone could ask for. Curtis, Chase, and Kenny at CU Boulder who never made me feel like I was too old to be a grad student. And my wife, Lindsay, who kept me focused on the light at the end of the tunnel.

\vskip 1in

\noindent
Evan Owen\\
March 22, 2023
\cleardoublepage


\begin{abstractpage}

I derive a formulation of the 2-dimensional critical Ising model on non-uniform simplicial lattices. Surprisingly, the derivation leads to a set of geometric constraints that a lattice must satisfy in order for the model to have a well-defined continuum limit. I perform Monte Carlo simulations of the critical Ising model on discretizations of several non-trivial manifolds including a twisted torus and a 2-sphere and I show that the simulations are in agreement with the 2d Ising CFT in the continuum limit. I discuss the inherent benefits of using non-uniform simplicial lattices to study quantum field theory and how the methods developed here can potentially be generalized for use with other theories.

\end{abstractpage}
\cleardoublepage


\tableofcontents
\cleardoublepage


\newpage
\listoffigures
\cleardoublepage

\chapter*{List of Abbreviations}


\begin{center}
  \begin{tabular}{lll}
    \hspace*{2em} & \hspace*{1in} & \hspace*{4.5in} \\
    CFT  & \dotfill & Conformal Field Theory \\
    DEC  & \dotfill & Discrete Exterior Calculus \\
    FEM  & \dotfill & Finite Element Method \\
    IR  & \dotfill & Infrared \\
    OPE  & \dotfill & Operator Product Expansion \\
    QCD  & \dotfill & Quantum Chromodynamics \\
    QFT  & \dotfill & Quantum Field Theory \\
    UV  & \dotfill & Ultraviolet \\
  \end{tabular}
\end{center}
\cleardoublepage


\newpage
\endofprelim
        
\cleardoublepage

\chapter{Introduction}
\label{chapter:intro}
\thispagestyle{myheadings}

In this thesis, I develop methods for simulating the critical Ising model on simplicial discretizations of 2-dimensional Riemannian surfaces. Prior to this work, almost all formulations of the Ising model were defined on uniform discretizations of a torus constructed from squares, rectangles, or equilateral triangles. By allowing for the use of non-uniform lattices, the new techniques developed here allow for simulations of the critical Ising model on a wider variety of 2-dimensional manifolds. I will describe how this additional flexibility is beneficial for the study of conformal field theories, and I will present numerical results for several examples of non-trivial manifolds which cannot be constructed from traditional toroidal lattices. In all of these examples the simulation results are in good agreement with exact solutions of the 2d Ising CFT in the continuum limit. Finally, I will discuss how the methods developed here can potentially be generalized for use with other theories.

In the remainder of this chapter, I discuss the historical context and the underlying scientific motivation for this work. In Chapter \ref{chapter:simplicial}, I derive the critical coupling values for the 2d Ising model defined on a general simplicial lattice as a function of the lattice geometry. The derivation leads to geometrical constraints that must be satisfied in order for the model to have a well-defined continuum limit. In Chapters \ref{chapter:affine} and \ref{chapter:sphere}, I use the result of this derivation to perform simulations of the critical Ising model on a uniform triangular lattice under an arbitrary affine transformation and a 2-sphere. Finally, in Chapter \ref{chapter:conclustion}, I make some concluding remarks about the significance of this work and potential future applications.

I aim for this thesis to be at the level of a graduate student in theoretical physics. This thesis is not meant to provide a thorough description of all of the mathematical and physical concepts that are employed. However, when I make use of less well-known concepts I will attempt to provide at least enough background to motivate the task at hand.

This work was supported by the U.S Department of Energy under Award No. DE-SC0019139.

\section{The Ising model and universality}

The Ising model was originally conceived by the German physicist Wilhelm Lenz in 1924 as a simple model of ferromagnetism. The 1-dimensional case was solved by Lenz' student Ernst Ising~\cite{Ising1925BeitragZT}, and the 2-dimensional case was later solved by Lars Onsager~\cite{Onsager1944CrystalSI}.

The traditional Ising model consists of a square lattice with a spin variable at each vertex which can take the values $\pm 1$ corresponding to spin up and down. Each spin interacts only with its nearest neighbors with an interaction strength which is inversely proportional to the temperature of the system. In the weak coupling limit (i.e. high temperature) the equilibrium state of the system is a disordered state where the spins are random, while the equilibrium state in the strong coupling limit (i.e. low temperature) is an ordered state where the majority of the spins are aligned in the same direction. This two-phase behavior is similar to that of a real ferromagnet.

In 2 or more dimensions there exists a critical temperature at which a continuous phase transition between the ordered and disordered phases occurs. At the critical point the system is characterized by a set of critical exponents which describe how the thermodynamic properties of the system behave at temperatures in the vicinity of the critical point. Many thermodynamic systems have such critical points, and somewhat surprisingly these systems can be grouped into ``universality classes'' which share the same set of critical exponents. As an example, the thermodynamic properties of water at its critical point are described by the same critical exponents as the 3-dimensional Ising model, even though these two physical systems seem to be completely unrelated.

\section{Quantum field theory and the lattice}

An important connection exists between the theory of critical phenomena and quantum field theory (QFT). By definition, quantum field theories are invariant under the Lorentz transformations of special relativity. In almost all cases, thermodynamic systems at critical points are described by a class of quantum field theories that are invariant under conformal transformations, which comprise the conformal group that includes the Lorentz group as a subgroup. These conformally invariant theories are known as conformal field theories (CFT) and are in one-to-one correspondence with the universality classes of critical phenomena. For example, the conformal field theory which describes the continuum limit of the 2d Ising model at its critical point is known as the 2d Ising CFT.

Although the original motivation for studying critical phenomena was to understand properties of materials at thermodynamic phase transitions, applications to many other branches of physics later appeared. In the context of this thesis, the most important of these applications relates to the use of the lattice as a tool in the study of quantum field theory. The lattice is important because it is currently the only well-established tool for modeling non-perturbative effects in QFT. Because of this, a great deal of effort has been invested in developing techniques for performing computational lattice simulations since the early 1980s~\cite{Creutz1980MonteCS}. In particular, the lattice is the primary tool used to study quantum chromodynamics (QCD), which is an inherently non-perturbative field theory at energies below the scale of quark confinement.

This brings us to the connection between lattice quantum field theory and the study of critical phenomena. Interactions in QFT are sensitive to the small-scale or ultraviolet (UV) description of a theory, which in the case of a lattice model is determined by the lattice spacing $a$. In order to properly describe a continuum quantum field theory, all lattice dependence of physical observables must vanish in the limit that the lattice spacing is taken to zero. To measure a physical quantity such as a mass $m$ with a corresponding dimensionless correlation length $\xi = m a$ on the lattice, it is necessary that the quantity $\xi/a$ remain finite as $a \to 0$, which is only possible if the correlation length diverges in the continuum limit. But indeed, it is a well known property of the study of critical phenomena that correlation lengths diverge at the critical points where phase transitions of thermodynamic systems occur. Therefore, in order to perform a valid lattice QFT simulation it is necessary to approach such a critical point in order for lattice discretization effects to vanish in the continuum limit. This constraint is in contrast to the study of classical (or non-interacting) field theories on a lattice, which do not have a strong sensitivity to the UV cutoff and therefore always have a valid continuum limit as the lattice spacing is taken to zero. In the context of the renormalization group in quantum field theory, these critical points are known as ``fixed points'' because they are the points where the parameters of the theory remain fixed as the UV cutoff is taken to infinity.

In the case of QCD and many other non-Abelian gauge theories, as the UV cutoff increases above the confinement scale the interaction strength goes exponentially to zero due to a phenomenon known as ``asymptotic freedom'' ~\cite{Gross1973UltravioletBO}. This occurs so long as the number of fermions in the theory is below a certain threshold. The correlation lengths of the operators in the resulting free theory are indeed divergent in the continuum limit, justifying the use of the lattice as a tool for studying such theories. This type of critical point is known as a UV fixed point because asymptotic freedom occurs in the high-energy (or ultraviolet) limit.

The critical point of the Ising model in 2 and 3 dimensions is fundamentally different from an asymptotically free theory because the interaction strength must be tuned to a precise nonzero value so that the correlation lengths of the theory diverge. This is known as an IR fixed point because the long-distance behavior of the theory is an emergent property determined by the relevant conformal field theory. Although the Ising model is relatively simple compared to a theory like QCD, both require the same theoretical framework to understand how the lattice theory approaches the continuum theory as the lattice spacing goes to zero. Thus, the Ising model is commonly used for developing new lattice techniques which can later be applied to more general problems in lattice field theory.

Finally, I will mention another interesting type of IR fixed point which occurs in non-Abelian gauge theories when the number of fermions in the theory lies in a specific range known as the ``conformal window''. These are known as a Banks-Zaks fixed points~\cite{Banks1982OnTP} and such theories are being actively studied on the lattice as candidates for describing a composite Higgs boson~\cite{Appelquist2020NearconformalDI,Appelquist2021GoldstoneBS}.


\section{Benefits of a spherical lattice}

In a conformal field theory, the energy spectrum of particles is continuous, with no lower limit on the mass of the lightest particle. This in contrast to a ``gapped'' theory like QCD, where the lightest stable particle has a nonzero mass $m$, and on a finite lattice of size $L$ finite volume effects are expected to decay like $\exp(-mL)$. Correlators in a CFT instead decay like a power law, which is much slower than the exponential decay of a gapped theory. This poses a major problem for numerical simulations on a finite lattice because the lattice volume (and the computational complexity) must grow exponentially to suppress finite volume effects. Even with modern computers, this results in a major limitation on how well an infinite-volume CFT can be modeled. A typical workaround is to parametrize the dependence of the theory on the size of the lattice. So-called finite size scaling methods can then be used to infer information about the infinite-volume theory~\cite{Fisher1972ScalingTF,Landau1976FinitesizeBO,Milchev1986FinitesizeSA}.

An alternative to these finite-size scaling methods that I will explore here is to exploit the properties of conformal symmetry in order to study the theory on manifolds other thant the periodic square lattices traditionally used for lattice simulations. In Appendix \ref{appendix:embedding_space} I describe how a CFT in flat space can be mapped onto other manifolds which are related to flat space via a Weyl transform. One useful choice is the ``cylinder'' of radial quantization, $\mathbb{R} \times S^{d-1}$, where the dilatation operator becomes the generator of translations in the radial direction. Instead of the power law behavior of flat space, correlation functions of CFT operators in radial quantization decay exponentially in the radial direction, much like correlators in a gapped theory. For a lattice simulation, this formulation therefore provides a powerful mitigation for finite volume effects. Another useful choice of manifold is the sphere $S^d$, which is a compact space and therefore has no finite volume effects at all. The main challenge with lattice simulations on these spherical manifolds is that it is very difficult to define a lattice theory on a discretized sphere in 2 or more dimensions.


Here, I describe a method for simulating the critical 2d Ising model on $S^2$ which results in restoration of conformal symmetries in the continuum limit. With this formulation, the properties of the 2d Ising CFT can be directly measured on the lattice and are in good agreement with the exact continuum values.

\section{Monte Carlo simulation details}

The research reported in this work made use of computing and long-term storage facilities of the USQCD Collaboration, which are funded by the Office of Science of the U.S. Department of Energy.

For all of our simulations, we use a combination of Metropolis~\cite{Metropolis1953EquationOS} and Wolff cluster~\cite{Wolff1989CollectiveMC} update algorithms to generate a Markov chain of statistically independent Ising spin configurations. We define statistically independent such that the longest autocorrelation time of the measurements that we perform is approximately $0.5$ in units of consecutively measured spin configurations. For each lattice size, we typically generate $\mathcal{O}(10^6)$ statistically independent spin configurations. To obtain the next configuration in the Markov chain, we perform a fixed number of Wolff updates so that the sum of the cluster sizes is at least as large as the lattice volume. We then perform a fixed number of Metropolis updates in order to satisfy the autocorrelation requirement above. Starting from either a completely random configuration or a completely uniform configuration (hot or cold start), we discard the first 100 configurations and then perform measurements on all subsequent configurations. We define the uncertainty in our reported measurements as the standard deviation of the mean value measured.

\cleardoublepage

\chapter{The simplicial Ising model}
\label{chapter:simplicial}
\thispagestyle{myheadings}

\graphicspath{{2_Simplicial/Figures/}}

As described in the introduction, lattice simulations on discretizations of a sphere can mitigate finite volume effects in conformal field theories. However, in 2 or more dimensions this is difficult because only a finite number of uniform discretizations of a sphere exist. This is related to the fact that the symmetry group of a $d$-sphere is O$(d+1)$, which only has a finite number of discrete subgroups for $d \geq 2$. For a 2d theory, the largest finite subgroup of O$(3)$ is the icosahedral point group with 120 group elements. Similarly, the largest finite subgroup of O$(4)$ is the Coxeter group $H_4$ with 14400 group elements. This means that for a uniform discretization of a sphere in more than 2 dimensions there is a fundamental minimum lattice spacing that can be obtained, obstructing any attempt to perform a reliable extrapolation to the continuum limit.

We could avoid this limitation by devising a method for simulating a theory on a non-uniform lattice which restores the continuum symmetries of the sphere as the effective lattice spacing goes to zero. In fact, this is exactly the approach that the finite element method (FEM) uses to solve classical field theories on arbitrary manifolds. FEM uses the geometry of a discretized Riemannian manifold to define discrete operators which converge to their continuum counterparts as the lattice spacing goes to zero.

Interacting quantum field theories are not so forgiving. Besides the method developed for the 2d Ising CFT in this thesis, there are no established methods for defining discrete operators for interacting quantum theories on non-uniform lattices which converge to their continuum counterparts. In fact, all of the currently established lattice methods for quantum theories are only valid for highly symmetric geometries like square and equilateral triangular lattices. With this motivation in mind, in this chapter I will derive a method for simulating the 2d Ising model on a non-uniform simplicial lattice.

\section{Kramers-Wannier duality}
\label{sec:kramers_wannier}

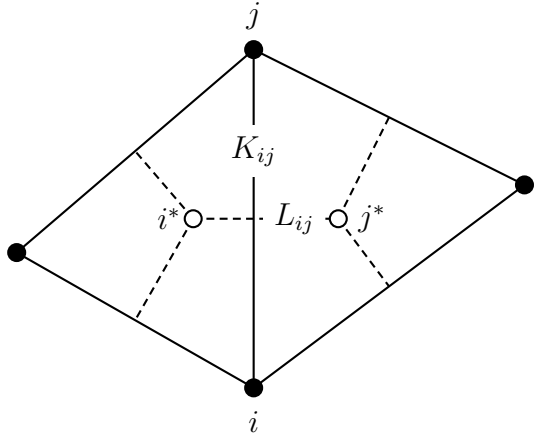
\begin{figure}
    \centering
    \begin{tikzpicture}[scale=4.5]

\coordinate (v1) at (0,0);
\coordinate (v2) at (0,1);
\coordinate (v3) at (-0.7, 0.4);
\coordinate (v4) at (0.8, 0.6);

\coordinate (e12) at (0, 0.5);
\coordinate (e13) at (-0.35, 0.2);
\coordinate (e14) at (0.4, 0.3);
\coordinate (e23) at (-0.35, 0.7);
\coordinate (e24) at (0.4, 0.8);

\coordinate (c1) at (-0.178571, 0.5);
\coordinate (c2) at (0.25, 0.5);

\coordinate (n13) at (-0.0545367, 0.717061);
\coordinate (n23) at (-0.0158736, 0.310186);
\coordinate (n123) at (-0.428571, 0.5);
\coordinate (n14) at (0.1, 0.7);
\coordinate (n24) at (0.138197, 0.276393);
\coordinate (n124) at (0.5,0.5);

\draw[thick] (v1) -- (v2);
\draw[thick] (v1) -- (v3);
\draw[thick] (v1) -- (v4);
\draw[thick] (v2) -- (v3);
\draw[thick] (v2) -- (v4);

\draw[thick, densely dashed] (c1) -- (c2);
\draw[thick, densely dashed] (c1) -- (e13);
\draw[thick, densely dashed] (c1) -- (e23);
\draw[thick, densely dashed] (c2) -- (e14);
\draw[thick, densely dashed] (c2) -- (e24);


\draw [thick,fill=black] (v1) circle[radius=0.025];
\draw [thick,fill=black] (v2) circle[radius=0.025];
\draw [thick,fill=black] (v3) circle[radius=0.025];
\draw [thick,fill=black] (v4) circle[radius=0.025];

\draw [thick,fill=white] (c1) circle[radius=0.025];
\draw [thick,fill=white] (c2) circle[radius=0.025];

\node at (0.0, -0.1) {$i$};
\node at (0.0, 1.1) {$j$};
\node at (-0.25, 0.5) {$i^*$};
\node at (0.35, 0.5) {$j^*$};

\node[fill=white] at (0,0.7) {$K_{ij}$};
\node[fill=white] at (0.12,0.5) {$L_{ij}$};



\end{tikzpicture}
    \caption{Couplings for a single edge on both the simplicial lattice (closed circles, solid lines) and its trivalent dual (open circles, dashed lines).}
    \label{fig:simplicial_couplings}
\end{figure}

We consider the Ising model defined on a 2d simplicial complex. A simplicial complex in 2 dimensions is defined by a set of vertices, edges, and triangular faces (i.e. 0-, 1-, and 2-simplices in the language of simplicial geometry) where each edge is shared by exactly two faces. At each vertex $i$ is a lattice site with a spin variable $\sigma_i = \pm 1$, corresponding to spin up or spin down. Each edge has a real-valued coupling constant $K_{ij}$. An example of a part of such a lattice is shown in Fig. \ref{fig:simplicial_couplings}. The Ising model action on this triangular lattice is
\begin{equation}
\label{eq:tri_action}
    S_{\text{tri}} = - \sum_{\langle ij \rangle} K_{ij} \sigma_i \sigma_j
\end{equation}
where $\langle ij \rangle$ denotes a sum over all edges in the simplicial complex. At this point the boundary conditions are arbitrary, and for the examples in the next two chapters we choose manifolds which have no boundary.

For any 2d simplicial lattice we can also define a trivalent dual lattice with dual sites corresponding to the triangular faces of the original lattice. The dual sites are connected by dual edges which are in one-to-one correspondence with the edges of the simplicial lattice. We define a coupling constant $L_{ij}$ for each dual edge. An Ising model action can be defined on the dual lattice as
\begin{equation}
\label{eq:dual_action}
    S_{\text{dual}} = - \sum_{\langle ij \rangle} L_{ij} \sigma_i \sigma_j
\end{equation}
where new $\langle ij \rangle$ denotes a sum over all dual edges. The pair of couplings $K_{ij}$ and $L_{ij}$ corresponds to a single edge and its dual, though the indices of the sites are necessarily different for the two lattices. In Fig. \ref{fig:simplicial_couplings}, they are denoted here as $i$ and $j$ for the simplicial lattice and $i^*$ and $j^*$ for the trivalent dual lattice.

For both the triangular lattice and its trivalent dual, the partition function of the system is
\begin{equation}
    Z = \sum_{\{\sigma_i\}=\pm 1} e^{-S}
\end{equation}
This is a well-defined statistical system and thus its dynamics can be calculated via standard methods such as Monte Carlo simulations.

At this point, we don't have enough information to determine the geometry of this system. The positions of the vertices, the lengths of the edges, and the locations of the dual sites within the triangular faces are all unknown quantities and may not be well-defined. Only if the model has a well-defined continuum limit, then the geometry of the system is an emergent property defined by the correlation functions of the resulting continuum theory. With that in mind, we will first consider the abstract graph-theoretical problem defined by the system, and later we will determine its emergent geometry in the continuum limit.

To understand how the triangular simplicial lattice and its trivalent dual lattice are related, we will perform a so-called low-temperature expansion of the triangular lattice partition function and a high-temperature expansion of the dual lattice partition function. Our goal is to show that up to an overall constant of proportionality the two partition functions are exactly equal. This is a standard procedure for generalized Ising models known as Kramers-Wannier duality~\cite{PhysRev.60.252,Wegner1971DualityIG}. In fact, it's not important which lattice gets the low-temperature expansion and which gets the high-temperature expansion, as long as we do one of each.

\begin{figure}
    \centering
    \includegraphics[width=0.5\textwidth]{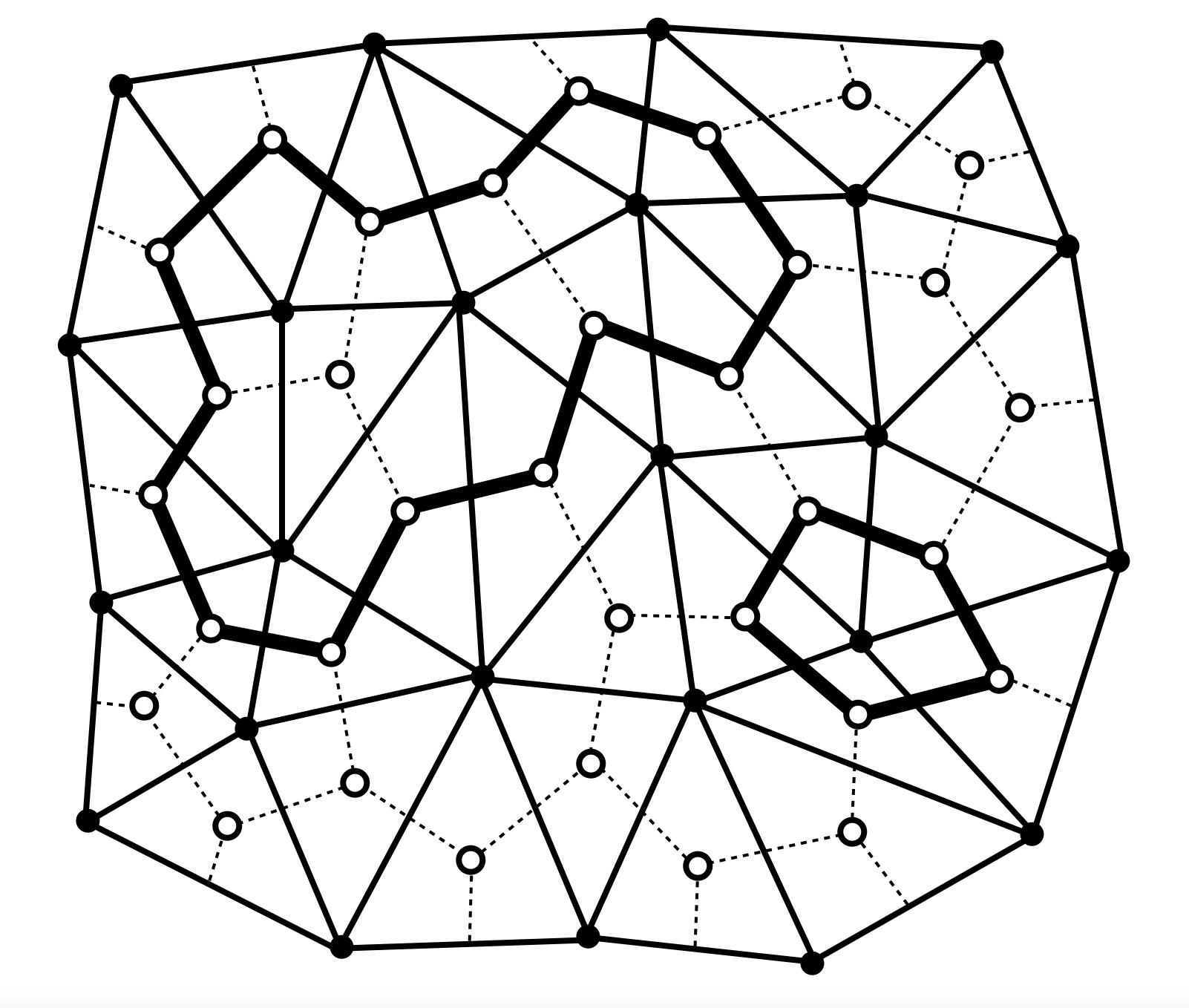}
    \caption{An example of a loop configuration for a non-uniform simplicial lattice and its trivalent dual lattice.}
    \label{fig:ising_loop}
\end{figure}

First, we perform the low-temperature expansion of the triangular lattice partition function. We would like to rewrite the sum over spin configurations in Eq. (\ref{eq:tri_action}) in terms of adjacent sites with unequal spins.
\begin{equation}
    S_{\text{tri}} = \prod_{\langle ij \rangle} e^{K_{ij}} \sum_{\{\sigma_i\}=\pm 1} \prod_{\sigma_i \neq \sigma_j} e^{-2K_{ij}}
\end{equation}
where the second product includes all pairs of adjacent sites in a given spin configuration with unequal spins. This is called a low-temperature expansion because the ``ground state'' is a state with no unequal spins, so the spins are either all up or all down as they are in a ferromagnet at very low temperatures. By picking out the edges with unequal spins, we split the lattice into regions with spin up and regions with spin down. The boundaries of these regions can be drawn as closed loops over the edges of the dual lattice. One example of a loop configuration $\Gamma$ is shown in Fig. \ref{fig:ising_loop}.
The sum over all spin configurations is then the set of unique configurations of closed loops on the dual lattice which we denote as $\{\Gamma\}$. There is also an additional irrelevant factor of 2 because each configuration has an equivalent configuration under the global transformation $\sigma_i \to - \sigma_i$. After discarding constant factors which do not affect the dynamics of the system, the triangular lattice partition function is
\begin{equation}
\label{eq:tri_kramers}
    Z_{\text{tri}} = \sum_{\{\Gamma\}} \prod_{\langle ij \rangle \in \Gamma} e^{-2 K_{ij}}
\end{equation}

Next, using the identity $e^{\pm x} = \cosh x \pm \sinh x$, the partition function of the dual lattice Ising model can be rewritten as
\begin{equation}
    Z_{\text{dual}} = \prod_{\langle ij \rangle} \cosh L_{ij} \sum_{\{\sigma_i\}=\pm 1} \prod_{\langle ij \rangle} (1 + \sigma_i \sigma_j \tanh L_{ij})
\end{equation}
Expanding out the second product will give a sum over all combinations of edges, i.e. $(\sigma_1 \sigma_2 \tanh L_{12})(\sigma_3 \sigma_4 \tanh L_{34})(\cdots)$. Because of the sum over spin configurations, these terms all vanish unless each site is included an even number of times, in which case all of the terms in the sum are equivalent. Because each site in a trivalent lattice has three neighbors, this means that the set of nonzero terms are those which include each site either twice or not at all. The graphical representation of this set is the same set of unique loop configurations $\{\Gamma\}$ that we used in the low-temperature expansion for the triangular lattice. Again discarding irrelevant constant factors, the partition function is
\begin{equation}
\label{eq:dual_kramers}
    Z_{\text{dual}} = \sum_{\{\Gamma\}} \prod_{\langle ij \rangle \in \Gamma} \tanh L_{ij}
\end{equation}
Each term in the sum is an average over all possible spin configurations which have equal Boltzmann weight in the limit that the temperature goes to infinity, justifying the term ``high-temperature'' expansion.

Although we use the term ``expansion'', both Eq. (\ref{eq:tri_kramers}) and (\ref{eq:dual_kramers}) are exact if we include all unique loop configurations admitted by the lattice. Comparing the two, we see that the partition functions of the two systems are equivalent if the coupling constants for each edge are related by
\begin{equation}
    e^{-2 K_{ij}} = \tanh L_{ij}
\end{equation}
which is equivalent to the symmetric form
\begin{equation}
\label{eq:dual_couplings}
    \sinh 2 K_{ij} \sinh 2 L_{ij} = 1
\end{equation}
This powerful relation holds for the Ising model on any simplicial lattice and its trivalent dual and will be used extensively in the remainder of this thesis.

\section{Wilson-Majorana fermion loop expansion}

In most treatments of the Ising model on triangular lattices, at this point authors introduce the star-triangle relation~\cite{Pokrovsky1982StartriangleRI}, which relates the partition functions of a uniform triangular lattice and its trivalent dual lattice in order to locate the critical point of the theory~\cite{Baxter1982ExactlySM}. However, the star-triangle method requires that each site in the triangular lattice must have exactly six neighbors and is therefore not applicable to the case of an arbitrary simplicial lattice. In fact, because the Euler characteristic of a sphere is nonzero it is impossible to construct a simplicial discretization of a sphere where all of the vertices have six neighbors. Instead of the star-triangle relation, here we will derive a new generalization of a fermionic method first introduced for an equilateral triangular lattice in~\cite{Wolff2020IsingMA} and generalized to the case of a triangular lattice under an affine transformation in~\cite{brower2022ising}. This will allow us to determine the critical values of the Ising model coupling constants for a non-uniform simplicial lattice.

\begin{figure}
    \centering
    \includegraphics[width=0.5\textwidth]{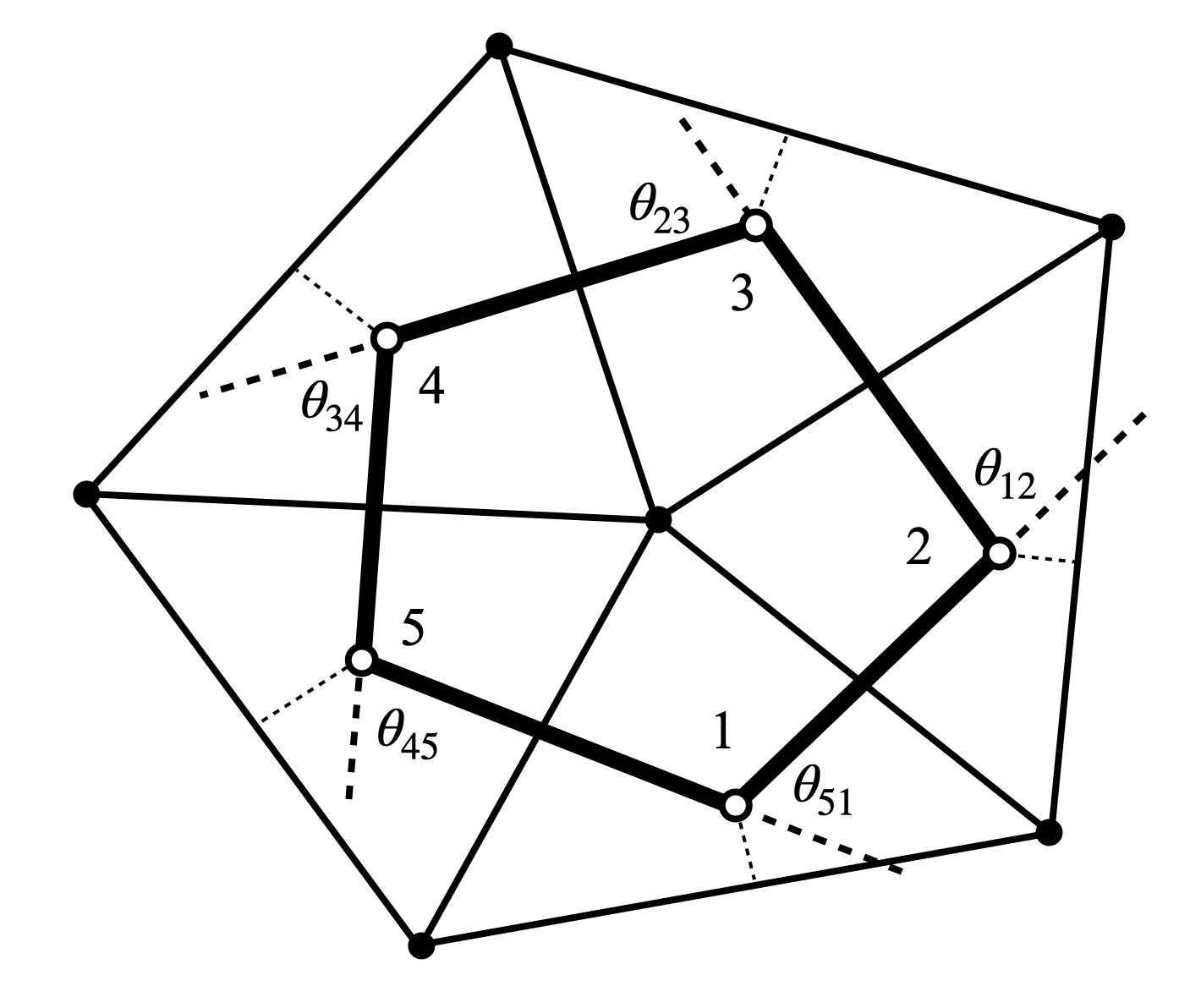}
    \caption{Phase angles around a loop which define the discrete spin connection for each edge.}
    \label{fig:phase_angles}
\end{figure}

To begin, we consider a hopping parameter expansion of a free Wilson-Majorana fermion defined on the trivalent dual lattice. The Wilson-Majorana fermion field $\psi_i^{\alpha}$ where $\alpha=1,2$ is a two-component spinor defined at each lattice site which obeys the charge conjugation constraint $\bar \psi_i^{\beta} = \psi^{\alpha}_i \epsilon^{\alpha \beta}$, i.e. $\bar \psi^1_i = -\psi^2_i$ and $\bar \psi^2_i = \psi^1_i$. The Wilson-Majorana action is
\begin{equation}
\label{eq:fermion_action}
    S_{\psi} = \dfrac{1}{2} \sum_i \bar \psi_i \psi_i - \sum_{\langle ij \rangle} \kappa_{ij} \bar \psi_i P_{ij} \psi_j
\end{equation}
where $\kappa_{ij}$ is a hopping parameter for each link and
\begin{equation}
    P_{ij} = \dfrac{1}{2} (1 - \hat e_{ij} \cdot \vec \sigma) \Omega_{ij}
\end{equation}
is a Wilson link matrix defined in terms of a unit vector $\hat e_{ij}$ and a spin connection $\Omega_{ij}$ from site $i$ to site $j$. We introduce the spin connection so that this action is valid even for a simplicial manifold with non-zero curvature, which will be important when we apply this formulation to the sphere. Because of this, some additional clarifications are in order. Following the conventions of \cite{Brower2016LatticeDF}, $\hat e_{ij}$ is defined as a unit vector from $i$ to $j$ in the plane of the triangle at the dual site $i$. This unit vector is not well-defined at this point in the derivation because we are still ignorant of the lattice geometry, but this will be remedied in the next section. For now we can simply assume that some valid geometry exists. The spin connection $\Omega_{ij}$ is discussed in great detail in \cite{Brower2016LatticeDF} and must be chosen to preserve the transformation properties of the spinor field under translations between sites. In general it depends on the choice of a local coordinate system at each site, but in the case of a 2-dimensional lattice we can simply define
\begin{equation}
    \Omega_{ij} = \pm e^{-i \theta_{ij} \sigma_z / 2}
\end{equation}
for some phase angle $\theta_{ij}$ for each link which is shown in Fig. \ref{fig:phase_angles} and will defined later. The sign of $\Omega_{ij}$ must be chosen such that the integrated curvature around each loop vanishes in the continuum limit, but this is always possible on any manifold which admits a spin structure.

We will now demonstrate that the fermion partition function
\begin{equation}
\label{eq:fermion_partition_fn}
    Z_{\psi} = \int \mathcal{D} \psi e^{-S_{\psi}}
\end{equation}
with Grassman path integral measure
\begin{equation}
    \mathcal{D} \psi \equiv \prod_i d \psi_i^1 d \psi_i^2
\end{equation}
is equivalent to the Ising partition function on the dual lattice (\ref{eq:dual_kramers}) up to an irrelevant constant. First, we expand the exponential and discard all terms above first order which vanish because they are Grassman-valued.
\begin{equation}
    Z_{\psi} = \int \mathcal{D} \psi_i \prod_i  \left( 1 - \dfrac{1}{2} \bar \psi_i \psi_i \right) \prod_{\langle ij \rangle} \left( 1 + \dfrac{1}{2} \bar \psi_i (1 - \hat e_{ij} \cdot \vec \sigma) \Omega_{ij} \psi_j \right)
\end{equation}
Now, following the standard rules for integration of Grassman variables, the only nonzero terms are those which include every pair $\bar \psi_i \psi_i$ exactly once. The set of nonzero terms is exactly described by the same set of unique loop configurations $\{\Gamma\}$ that we used in Sec. \ref{sec:kramers_wannier}. Note that it does not matter which direction we choose to traverse each loop, the result will be the same either way.

Again discarding irrelevant constant factors, the fermion partition function becomes
\begin{equation}
    Z_{\psi} = \sum_{\{\Gamma\}} (-1)^n \operatorname{tr} \left[ \prod_{\langle ij \rangle \in \Gamma} \dfrac{1}{2} \kappa_{ij} (1 - \hat e_{ij} \cdot \vec \sigma) \Omega_{ij} \right]
\end{equation}
where $n$ is the number of closed loops in $\Gamma$. Now, for every edge $ij$ in any loop we can always choose a local coordinate system at each site such that
\begin{equation}
    \hat e_{ij} \cdot \vec \sigma = -\sigma_x =
    \begin{pmatrix}
        0 & -1 \\
        -1 & 0
    \end{pmatrix}
\end{equation}
Again following the conventions of \cite{Brower2016LatticeDF}, the phase angle $\theta_{ij}$ is the angle of rotation encountered at site $j$ when traversing a loop as shown in Fig. \ref{fig:phase_angles}. For any manifold that admits a spin structure, the sign factors in $\Omega_{ij}$ can always be chosen so that every closed loop has a phase factor of $-1$ to cancel the minus signs out front.

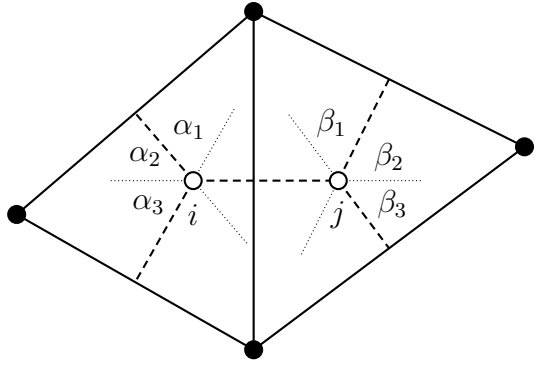
\begin{figure}
    \centering
    \begin{tikzpicture}[scale=4.5]

\coordinate (v1) at (0,0);
\coordinate (v2) at (0,1);
\coordinate (v3) at (-0.7, 0.4);
\coordinate (v4) at (0.8, 0.6);

\coordinate (e12) at (0, 0.5);
\coordinate (e13) at (-0.35, 0.2);
\coordinate (e14) at (0.4, 0.3);
\coordinate (e23) at (-0.35, 0.7);
\coordinate (e24) at (0.4, 0.8);

\coordinate (c1) at (-0.178571, 0.5);
\coordinate (c2) at (0.25, 0.5);

\coordinate (n13) at (-0.0545367, 0.717061);
\coordinate (n23) at (-0.0158736, 0.310186);
\coordinate (n123) at (-0.428571, 0.5);
\coordinate (n14) at (0.1, 0.7);
\coordinate (n24) at (0.138197, 0.276393);
\coordinate (n124) at (0.5,0.5);

\draw[thick] (v1) -- (v2);
\draw[thick] (v1) -- (v3);
\draw[thick] (v1) -- (v4);
\draw[thick] (v2) -- (v3);
\draw[thick] (v2) -- (v4);

\draw[thick, densely dashed] (c1) -- (c2);
\draw[thick, densely dashed] (c1) -- (e13);
\draw[thick, densely dashed] (c1) -- (e23);
\draw[thick, densely dashed] (c2) -- (e14);
\draw[thick, densely dashed] (c2) -- (e24);

\draw[densely dotted] (c1) -- (n13);
\draw[densely dotted] (c1) -- (n23);
\draw[densely dotted] (c1) -- (n123);
\draw[densely dotted] (c2) -- (n14);
\draw[densely dotted] (c2) -- (n24);
\draw[densely dotted] (c2) -- (n124);

\draw [thick,fill=black] (v1) circle[radius=0.025];
\draw [thick,fill=black] (v2) circle[radius=0.025];
\draw [thick,fill=black] (v3) circle[radius=0.025];
\draw [thick,fill=black] (v4) circle[radius=0.025];

\draw [thick,fill=white] (c1) circle[radius=0.025];
\draw [thick,fill=white] (c2) circle[radius=0.025];

\node at (-0.178571, 0.4) {$i$};
\node at (0.25, 0.4) {$j$};


\node at (-0.19, 0.65) {$\alpha_{1}$};
\node at (-0.32, 0.57) {$\alpha_{2}$};
\node at (-0.31, 0.43) {$\alpha_{3}$};

\node at (0.23, 0.67) {$\beta_{1}$};
\node at (0.40, 0.57) {$\beta_{2}$};
\node at (0.41, 0.43) {$\beta_{3}$};

\end{tikzpicture}
    \caption{Definition of the 6 angles used in Eq. (\ref{eq:fermion_ising_match}). The open circles are the circumcenters of the two triangles.}
    \label{fig:generic_ising_angles}
\end{figure}

For each link we have a factor
\begin{equation}
    \dfrac{1}{2}\kappa_{ij}
    \begin{pmatrix}
        1 & 1 \\ 1 & 1
    \end{pmatrix} 
    \begin{pmatrix}
        e^{- i \theta_{ij} / 2} & 0 \\
        0 & e^{ i \theta_{ij} / 2}
    \end{pmatrix} = \dfrac{1}{2}
    \begin{pmatrix}
        1 \\ 1
    \end{pmatrix}
    \begin{pmatrix}
        e^{- i \theta_{ij} / 2} & e^{i \theta_{ij} / 2}
    \end{pmatrix}
\end{equation}
Simplifying the products of these factors around each loop and evaluating the trace, the partition function then becomes
\begin{equation}
\label{eq:fermion_loop}
    Z_{\psi} = \sum_{\{\Gamma\}} \prod_{\langle ij \rangle \in \Gamma} \kappa_{ij} \cos \dfrac{\theta_{ij}}{2} \;.
\end{equation}
We can't immediately relate this to the dual lattice Ising model partition function because the angles here are defined at the vertices of each loop, whereas the Ising coupling coefficients are defined on the edges. However, expanding out all possible loops and matching terms in (\ref{eq:dual_kramers}) and (\ref{eq:fermion_loop}) we find that they are exactly equal if
\begin{equation}
\label{eq:fermion_ising_match}
    \tanh^2 L_{ij} = \kappa_{ij}^2 \dfrac{\cos (\alpha_{2}/2) \cos (\alpha_{3}/2) \cos (\beta_{2}/2) \cos (\beta_{3}/2)}{\cos (\alpha_{1}/2) \cos (\beta_{1}/2)}
\end{equation}
where the six angles are defined in Fig. \ref{fig:generic_ising_angles}. We can use Eq. (\ref{eq:dual_couplings}) to relate the triangular lattice coupling $K_{ij}$ to the corresponding dual lattice coupling $L_{ij}$, and by association we now have three different models with equivalent partition functions.

\section{Lattice continuum limit}

We have now shown that the Ising model on both the simplicial lattice and its trivalent dual and the free Wilson-Majorana fermion all have equivalent partition functions, and therefore they all describe systems with equivalent dynamics. But we haven't yet identified the critical point of the system, nor have we determined the appropriate geometry of the continuum theory (assuming such a theory exists). To do this, we will show that the lattice fermion action (\ref{eq:fermion_action}) converges to its continuum counterpart.

We expand (\ref{eq:fermion_action}) in the effective lattice spacing $a$ using
\begin{equation}
    \psi_j = \psi_i + \vec l_{ij}^* \cdot \vec \nabla \psi_i + \mathcal{O}(a^2)
\end{equation}
and
\begin{equation}
    \Omega_{ij} = e^{- i \vec l_{ij}^* \cdot \vec \omega} = 1 - i \vec l_{ij}^* \cdot \vec \omega + \mathcal{O}(a^2)
\end{equation}
where $\vec l_{ij}^*$ is a vector from $i$ to $j$ with length $l_{ij}^* = \mathcal{O}(a)$ and $\vec \omega$ is the continuum spin connection. The action becomes
\begin{equation}
\label{eq:fermion_action_expand}
\begin{split}
    S_{\psi} &= \dfrac{1}{2} \sum_i \bar \psi_i \psi_i - \dfrac{1}{2} \sum_{\langle ij \rangle} \kappa_{ij} \bar \psi_i (1 - \hat e_{ij} \cdot \vec \sigma) (1 - i \vec l_{ij}^* \cdot \vec \omega) (1 + \vec l_{ij}^* \cdot \vec \nabla ) \psi_i + \mathcal{O}(a^2) \\
    &= \dfrac{1}{2} \sum_i \left( 1 - \dfrac{1}{2} \sum_{j \in \langle ij \rangle} \kappa_{ij} \right) \bar \psi_i \psi_i + \dfrac{1}{2} \sum_{\langle ij \rangle} \kappa_{ij} \bar \psi_i (\hat e_{ij} \cdot \vec \sigma) (\vec l_{ij}^* \cdot \vec D) \psi_i
\end{split}
\end{equation}
where $\vec D = \vec \nabla - i \vec \omega$ is the covariant derivative for the spinor field and in the second line we have used $\bar \psi_i \vec \sigma \psi_i = 0$ and discarded all terms of $\mathcal{O}(a^2)$.

We would like to show that this is equivalent to the continuum action for a free fermion on a Riemannian manifold,
\begin{equation}
\label{eq:fermion_action_cont}
    S_{\text{cont}} = \dfrac{1}{2} \int d^2 x~ \sqrt{g} \bar \psi(\vec x) (m + \vec \sigma \cdot \vec D) \psi (\vec x) \;.
\end{equation}
In order to get Eq. (\ref{eq:fermion_action_expand}) into this form, we need the $\vec \sigma$ and $\vec D$ in the lattice action to be contracted, which is not possible for an arbitrary choice of lattice vectors in a non-uniform lattice. However, provided that all adjacent triangles have the same circumradius, the following identity is true for a lattice which is the circumcenter dual of a non-uniform simplicial lattice in 2 dimensions:
\begin{equation}
    \sum_{j \in \langle ij \rangle} l_{ij}^{\mu} l_{ij}^{*\nu} = 2 A_i \epsilon^{\mu \nu}
\end{equation}
where $\vec l_{ij}$ is the lattice vector in the simplicial lattice dual to $\vec l^*_{ij}$ and $A_i$ is the area of the triangular face dual to site $i$.

In order to apply this identity, it is necessary for $\kappa_{ij}$ to be proportional to the corresponding triangular lattice length $l_{ij}$.  Substituting this all into the action we obtain
\begin{equation}
    S_{\psi} = \dfrac{1}{2} \sum_i A_i \bar \psi_i \left(m_i + \vec \sigma \cdot \vec D \right) \psi_i + \mathcal{O}(a^2)
\end{equation}
with a dimensionless mass parameter
\begin{equation}
\label{eq:crit_fermion_mass}
    m_i = \dfrac{1}{A_i} \left( 1 - \dfrac{1}{2} \sum_{j \in \langle ij \rangle} \kappa_{ij} \right) \;.
\end{equation}
In the limit $a \to 0$ we identify the limit of the discrete integration measure $A_i \to d^2 x \sqrt{g}$ and find that the lattice action converges to the continuum action in Eq. (\ref{eq:fermion_action_cont}), as required. Additionally, we can identify the critical point as the massless limit of the free fermion theory by setting $m_i=0$. This implies that the critical hopping parameter values for each link are defined by
\begin{equation}
\label{eq:kappa_crit}
    \kappa_{ij} = \dfrac{2 l_{ij}}{\sum_{k \in \langle ik \rangle} l_{ik}}
\end{equation}
Now, in order to have a well-defined critical point we must satisfy this relation for every link on the lattice simultaneously, but upon inspection we find that this is only possible if all of the triangular faces have equal perimeter. Combining Eq. (\ref{eq:fermion_ising_match}) and (\ref{eq:kappa_crit}) we obtain an expression which relates the dual lattice Ising coupling $L_{ij}$ to the geometry of the lattice
\begin{equation}
\label{eq:crit_couplings}
    \tanh^2 L_{ij} = \dfrac{4 l_{ij}^2 \cos (\alpha_{2}/2) \cos (\alpha_{3}/2) \cos (\beta_{2}/2) \cos (\beta_{3}/2)}{P_{\triangle}^2 \cos (\alpha_{1}/2) \cos (\beta_{1}/2)}
\end{equation}
where $P_{\triangle}$ is the triangle perimeter. We can also recover the critical couplings $K_{ij}$ for the triangular lattice through Eq. (\ref{eq:dual_couplings}).

The two geometrical constraints of equal triangle circumradius and equal triangle perimeter place strict limitations on the set of simplicial lattices for which it is possible to define a critical theory with a well-defined continuum limit. In Chapter \ref{chapter:affine} we will test the formulation derived here by simulating the critical Ising model on simplicial lattices constructed from identical skew triangles. Then in Chapter \ref{chapter:sphere} we will show that it is possible to construct non-uniform lattices which satisfy these geometric constraints in the continuum limit, allowing us to perform simulations of the critical Ising model on discretizations of the 2-sphere.

\cleardoublepage

\chapter{Ising model on an affine plane}
\label{chapter:affine}
\thispagestyle{myheadings}

\graphicspath{{3_Affine/Figures/}}

In this chapter I will test the formalism developed in the previous chapter by performing simulations of the critical Ising model on a uniform triangular lattice under an arbitrary affine transformation. The simulation results, first presented in~\cite{brower2022ising}, are in good agreement with the exact solution of the 2d Ising CFT on a torus with twisted boundary conditions. In addition, I will show that the simulation results are in good agreement with the infinite volume theory via finite-size scaling methods. Note that because here we are using a uniform lattice, the geometric constraints of uniform triangle circumradius and perimeter are automatically satisfied.

\section{Uniform simplicial Ising model}

\begin{figure}[h]
\begin{center}
\begin{tikzpicture}[>=stealth,scale=1.25]
    
    \foreach \y in {0,1,2,3}
    \foreach \x in {0,1,2,3} {
    \begin{scope}[shift={(\x+\y*0.5,\y*0.866)}]
    \draw[thick] (0,0) -- (1,0) -- (0.5,0.866) -- cycle;
    \draw[densely dashed] (0.25,0.433) -- (0.5,0.289) -- (0.5,0);
    \draw[densely dashed] (1,0.866) -- (1,0.866-0.289) -- (1.25,0.433);
    \draw[densely dashed] (0.5,0.289) -- (1,0.866-0.289);
    \draw (0.5,0.289) circle[radius=0.075];
    \draw (1,0.866-0.289) circle[radius=0.075];
    \end{scope}
    }
    \draw[thick] (2,4*0.866) -- (6,4*0.866) -- (4,0);

    \foreach \y in {0,1,2,3,4}
    \foreach \x in {0,1,2,3,4} {
    \begin{scope}[shift={(\x+\y*0.5,\y*0.866)}]
    \fill (0,0) circle[radius=0.075];
    \end{scope}
    }
    
    \begin{scope}[shift={(-1,2)}]
    \draw[thick,->] (0,0) -- (1,0);
    \draw[thick,->] (0,0) -- (-0.5,-0.866);
    \draw[thick,->] (0,0) -- (-0.5,0.866);
    
    \node[anchor=south] at (0.65,0.0) {$\hat{e}_1$};
    \node[anchor=south west] at (-0.25,0.5*0.866) {$\hat{e}_2$};
    \node[anchor=north west] at (-0.25,-0.5*0.866) {$\hat{e}_3$};
    \end{scope}
    
\end{tikzpicture}
\begin{tikzpicture}[scale=1.25]


\draw[thick] (-0.433,0) -- (1.732+0.433,0);
\draw[thick] (0.433,1.5) -- (2.598+0.433,1.5);
\draw[thick] (-0.433/2,-0.375) -- (0.866*1.25,1.5+0.375);
\draw[thick] (1.732+0.433/2,-0.375) -- (0.866*0.75,1.5+0.375);
\draw[thick] (1.732-0.433/2,-0.375) -- (2.598+0.433/2,1.5+0.375);
\draw[thick] (2.598-0.433/2,1.5+0.375) -- (2.598+0.433/2,1.5-0.375);
\draw[thick] (0.433/2,-0.375) -- (-0.433/2,0.375);

\fill (0,0) circle[radius=0.0833];
\fill (0.866,1.5) circle[radius=0.0833];
\fill (1.732,0) circle[radius=0.0833];
\fill (2.598,1.5) circle[radius=0.0833];

\node[anchor=south] at (0.866,0.0) {$K_1$};
\node[anchor=south] at (1.732,1.5) {$K_1$};
\node[anchor=south east] at (0.5,0.65) {$K_3$};
\node[anchor=north west] at (2.1,0.9) {$K_3$};
\node[anchor=south west] at (1.25,0.65) {$K_2$};

\end{tikzpicture}
\caption{On the left, the triangular graph (black dots, solid lines) and its hexagonal dual graph (open circles, dashed lines). On the right, the 3 independent couplings $K_i$ assigned to each triangle. The dual couplings $L_i$ are assigned to the corresponding dual links. This graph representation of the model does not show the emergent lattice geometry.}
\label{fig:graphsDuality}
\end{center}
\end{figure}
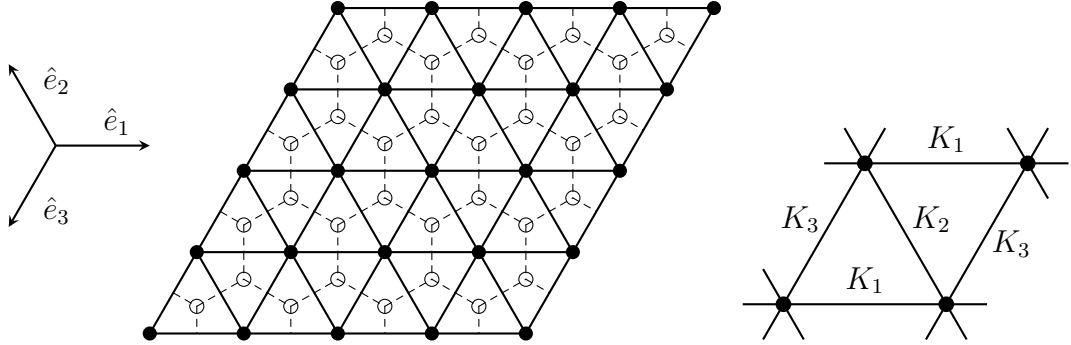

We consider the Ising model defined on a uniform triangular lattice and its hexagonal dual lattice, as depicted in Fig. \ref{fig:graphsDuality}. For each model we choose only three different coupling values, one for each lattice direction, denoted as $\{K_1, K_2, K_3\}$ and $\{L_1, L_2, L_3\}$ for the triangular and hexagonal lattices, respectively. Though the indexing conventions are different, this is a subset of the general set of simplicial lattices used in the derivation in Chapter \ref{chapter:simplicial} which preserve discrete translation symmetries in each lattice direction and rotation symmetry by 180°.

Eq. (\ref{eq:dual_couplings}) and (\ref{eq:crit_couplings}) give a one-to-one correspondence between the triangular lattice couplings $K_i$, the hexagonal couplings $L_i$, and the lattice geometry. Because the three coupling values on each triangle are equal, this guarantees that the geometry will also be uniform, i.e. the coupling values $K_i$ are in one-to-one correspondence with the triangle edge lengths $l_i$ and similarly for the trivalent dual lattice. This last point ensures that the triangles all have equal circumradius and perimeter, which ensures that the geometric constraints we obtained in the derivation in Chapter \ref{chapter:simplicial} will be satisfied. In fact, combining Eq. (\ref{eq:dual_couplings}) and (\ref{eq:crit_couplings}) for this specific case of a uniform triangular lattice results in the following remarkably simple relation between critical couplings and edge lengths
\begin{equation}
\label{eq:uniform_couplings}
    \sinh 2 K_i = 1 / \sinh 2 L_i = \dfrac{l_i^*}{l_i}
\end{equation}
where $l_i$ and $l_i^*$ are the corresponding edge lengths for the triangular and dual lattice, respectively.

\section{Modular parameter}
\label{sec:Modular}

\begin{figure}[h]
    \centering
    \begin{tikzpicture}[>=stealth,scale=2.0]
    
    \draw[->] (-1.2,0) -- (1.2,0);
    \draw[->] (0,-0.3) -- (0,2.0);
    
    \draw[loosely dashed] (1,0) arc (0:180:1);
    \draw[loosely dashed] (0.5,0) -- (0.5,0.866025);
    \draw[loosely dashed] (-0.5,0) -- (-0.5,0.866025);
    \draw[thick] (0.5,0.866025) arc (60:120:1);
    \draw[thick] (0.5,0.866025) -- (0.5,1.8);
    \draw[thick] (-0.5,0.866025) -- (-0.5,1.8);
    
    \node[anchor=west] at (1.3,0) {$\operatorname{Re} \tau$};
    \node[anchor=south] at (0,2.1) {$\operatorname{Im} \tau$};
    \draw (1,0) -- (1,-0.05);
    \draw (-1,0) -- (-1,-0.05);
    \node[anchor=north] at (1,-0.1) {$1$};
    \node[anchor=north] at (-1,-0.1) {$-1$};
    
    \begin{scope}
    \clip (0.5,0)--(0.5,1.8)--(-0.5,1.8)--(-0.5,0)--cycle;
    \fill[fill=black,opacity=0.1] (-0.5,0) rectangle (0.5,2) (0,0) circle[radius=1];
    \end{scope}
    
    \fill (0.5,0.866025) circle[radius=0.03125];
    \fill (0,1) circle[radius=0.03125];
    \fill (0.15625,1.2402) circle[radius=0.03125];
    \node[anchor=south west] at (0.57,0.85) {$\tau_{111} = e^{i \pi / 3}$};
    
    \node[anchor=south east] at (-0.65,1.15) {$\tau_{\Box} = i$};
    \draw[->] (-0.65,1.25) -- (-0.08,1.04);

    \node[anchor=south west] at (0.6,1.3) {$\tau_{456} = \dfrac{5}{32} \left(1 + 3 \sqrt{7} i \right)$};
    \draw[->] (0.6,1.45) -- (0.23,1.28);
    
    \draw[very thick,densely dashed] (0,0) -- (0.15625,1.2402) -- (1,0) -- cycle;
    
    \path(-2,0);
    \path(2,0);
    
    \end{tikzpicture}
    \caption{The fundamental domain of the modular parameter $\tau$ showing several explicit values.}
    \label{fig:fundamental_domain}
\end{figure}
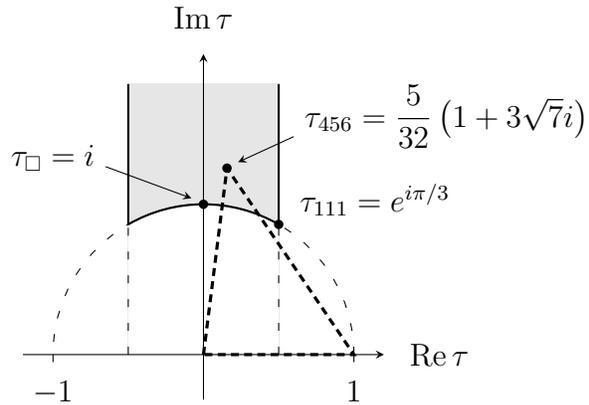

The first numerical test of our formalism is to embrace the periodic and finite nature of traditional lattice simulations and compare our model to the 2d Ising CFT defined on a torus with arbitrary modular parameter $\tau$. The modular parameter is a concept familiar to string theorists and is used to parameterize all of the possible boundary conditions of a torus. Put simply, if the torus is thought of as a tube, the modular parameter is a complex number which parameterizes how the tube is stretched and twisted before its two ends are glued together. The shaded region $\{\tau : |\tau| \ge 1, |\operatorname{Re} \tau| \le 1/2, \operatorname{Im} \tau > 0 \}$ shown in Fig. \ref{fig:fundamental_domain} indicates the fundamental domain of the modular parameter $\tau$. Each value of $\tau$ in this region defines a triangle from which we can construct a unique 2d lattice with periodic boundary conditions and the topology of a torus. We have indicated the locations of the equilateral case $(\tau_{111})$, the square case $(\tau_{\Box})$, and a skew triangle $(\tau_{456})$ which we will use as a representative example to perform our simulations. The heavy dashed lined is the triangle defined by $\tau_{456}$. The notation $\tau_{ijk}$ indicates that the triangle side lengths are proportional to $\{i,j,k\}$.

\begin{figure}
    \centering
    \begin{tikzpicture}[scale=1.5]
    
    \foreach \y in {0,1,2,3}
    \foreach \x in {0,1,2,3} {
    \begin{scope}[shift={(\x+\y*0.15625,\y*1.2402)}]
    \draw[thick] (0,0) -- (1,0) -- (0.15625,1.2402) -- cycle;
    \draw[densely dashed] (0.5*0.15625,0.5*1.2402) -- (0.5,0.57) -- (0.5,0);
    \draw[densely dashed] (1+0.5*0.15625,0.5*1.2402) -- (0.5+0.15625,1.2402-0.57) -- (0.5+0.15625,1.2402);
    \draw[densely dashed] (0.5,0.57) -- (0.5+0.15625,1.2402-0.57);
    \draw (0.5,0.57) circle[radius=0.05];
    \draw (0.5+0.15625,1.2402-0.57) circle[radius=0.05];
    \end{scope}
    }
    \draw[thick] (0.625,4*1.2402) -- (4+0.625,4*1.2402) -- (4,0);
    
    \foreach \y in {0,1,2,3,4}
    \foreach \x in {0,1,2,3,4} {
    \begin{scope}[shift={(\x+\y*0.15625,\y*1.2402)}]
    \fill (0,0) circle[radius=0.05];
    \end{scope}
    }

    \end{tikzpicture}
    \caption{The solid lines and closed circles indicate the triangular lattice defined by $\tau_{456}$ with $L=4$. The dashed lines and open circles indicate the dual hexagonal lattice with sites at the circumcenter of each triangle.}
    \label{fig:dual_lattice}
\end{figure}
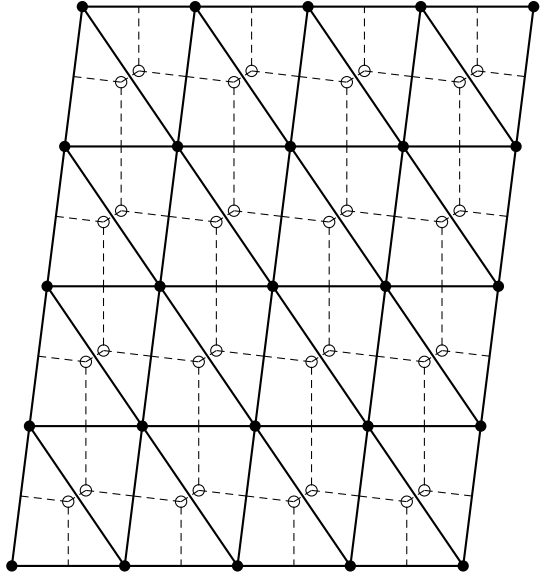

Without loss of generality, we can sort the triangular lattice lengths so that $l_1 \leq l_2 \leq l_3$. Then the modular parameter in the fundamental domain is
\begin{equation}\label{eq:mod_param}
    |\tau| = \dfrac{l_2}{l_1}, \qquad \arg(\tau) =  \cos^{-1}(-\hat{e}_1 \cdot \hat{e}_2) \; .
\end{equation}
where $\hat e_i = \vec l_i / l_i$ is a lattice basis vector in the $i$ direction. The lattice is implemented as a refined parallelogram with triangular cells with the appropriate side lengths as shown in Fig. \ref{fig:dual_lattice}.

It is interesting to note that the two parameters of the complex number $\tau$ describe a triangle under all affine transformations mod scaling. This is a fundamental property of simplicial geometry and also generalizes to higher dimensional simplices under the appropriate affine transformations.

The continuum two-point function for the critical Ising model on a torus with modular parameter $\tau$ is known to be~\cite{Francesco1987CriticalIC}
\begin{equation}
\label{eq:torus_exact}
    \langle \sigma(0) \sigma(z) \rangle = \left| \dfrac{\vartheta_1'(0 | \tau)}{\vartheta_1(z | \tau)} \right|^{1/4} \dfrac{\sum_{\nu=1}^4 |\vartheta_{\nu}(z/2|\tau)|}{\sum_{\nu=2}^4 |\vartheta_{\nu}(0|\tau)|}
\end{equation}
where $\vartheta_{\nu}(z|\tau)$ are the Jacobi theta functions and $z = x + iy$ is the separation vector in complex coordinates.

\begin{figure}[h]
    \centering
    \includegraphics{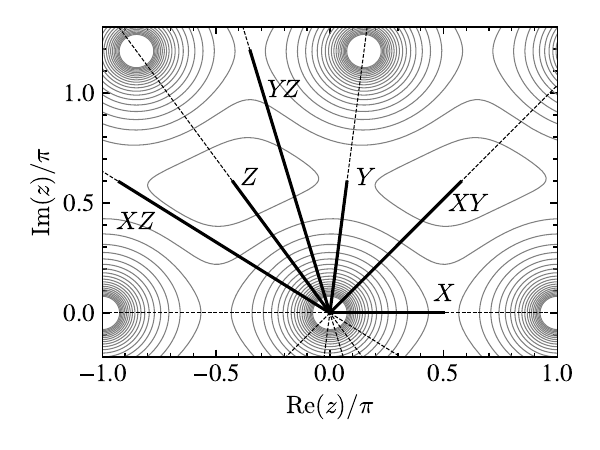}
    \caption{Contour plot of the continuum two-point function $\langle \sigma(z) \sigma(0) \rangle$ in the complex plane with modular parameter $\tau_{456}$, highlighting six axes along which we can easily measure the two-point function on the lattice. }
    \label{fig:torus_2pt_contour}
\end{figure}

In Fig. \ref{fig:torus_2pt_contour} we show a contour plot of the continuum 2d Ising spin-spin two-point function on a torus with modular parameter given by Eq.~\ref{eq:mod_param} for a triangle with side lengths $l_i \propto \{ 4,5,6 \}$ and critical coupling coefficients $K_i \simeq \{ 0.48648, 0.31824, 0.062829 \}$ computed using Eq. (\ref{eq:uniform_couplings}). On a triangular lattice, it is convenient to measure the two-point function along the six axes shown. It is sufficient to measure only the bold part of each axis due to the periodicity of the torus.

\begin{figure}
    \centering
    \includegraphics[width=0.49\textwidth]{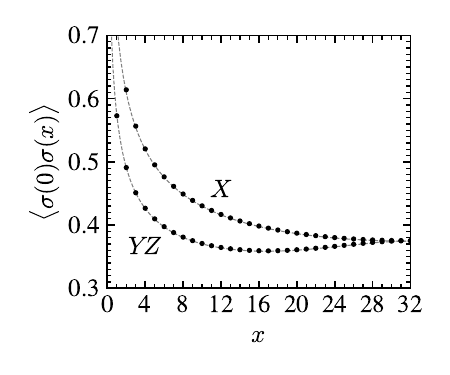}
    \includegraphics[width=0.49\textwidth]{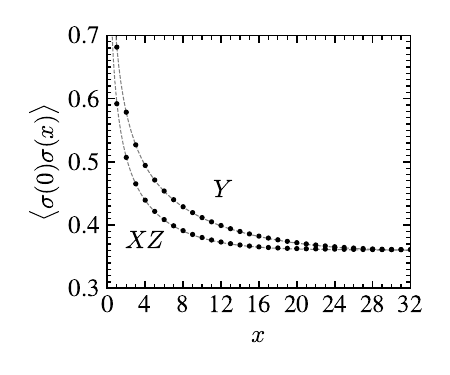}
    \includegraphics[width=0.49\textwidth]{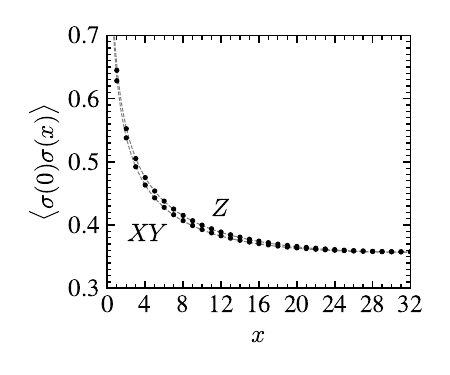}
    \caption{Two-point correlation function measured along the six axes shown in Fig. \ref{fig:torus_2pt_contour} for a triangular lattice with $l_i \propto \{ 4,5,6 \}$ and $L=64$. The gray lines are the exact correlation function from Eq. \ref{eq:torus_exact} and the black points are lattice data. The error bars are too small to be seen here.}
    \label{fig:ising_flat_crit}
\end{figure}

We perform a simultaneous fit to Eq. (\ref{eq:torus_exact}) using lattice measurements for all six of these axes. The only fit parameter is a single normalization factor. We can see in Fig. \ref{fig:ising_flat_crit} that the lattice data for all six axes is in excellent agreement with the continuum result, indicating that the emergent geometry of the lattice theory is consistent with the 2d Ising CFT.

\section{Finite-size scaling analysis}

In this section, we use the same lattice construction as in Sec.~\ref{sec:Modular}, but we now extract information about the continuum theory on an infinite plane via finite-size scaling. A similar analysis was done for an equilateral triangular lattice in \cite{Luo2008CriticalBO}. Our formalism allows us to construct the lattice from triangles with any side lengths. Here, we again use triangles with $l_i \propto \{ 4,5,6 \}$ because it is sufficiently different from an equilateral, isosceles, or right triangle and will show a clear difference between measurements along different triangle axes.

\begin{figure}[h]
    \centering
    \includegraphics{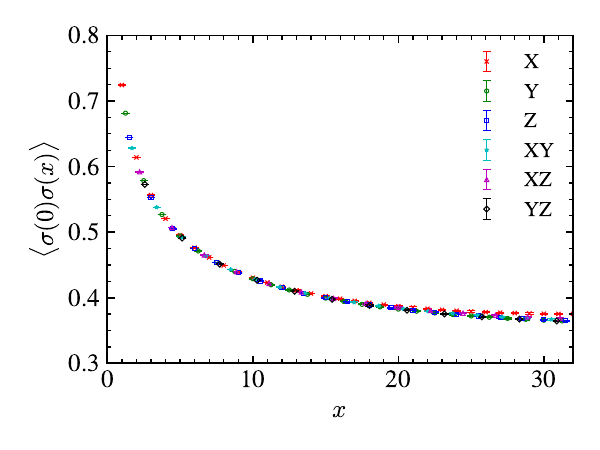}
    \caption{Spin-spin correlation function measured along each of the axes shown in Fig. \ref{fig:torus_2pt_contour} with distances scaled appropriately based on the lattice lengths $l_i$. Shown here for $l_i \propto \{ 4,5,6 \}$.}
    \label{fig:ising_r2_crit}
\end{figure}

Again using critical couplings computed using Eq. (\ref{eq:uniform_couplings}), we measure the spin-spin correlation
function along the six axes shown in
Fig. \ref{fig:torus_2pt_contour}. After scaling the step length for
each axis appropriately, the correlation functions collapse onto a single curve
at small separation, as shown in Fig. \ref{fig:ising_r2_crit}. At
large separation, wraparound effects cause the correlation function behavior to
depend slightly on which axis is being measured.

\begin{figure}[h]
    \centering
    \includegraphics{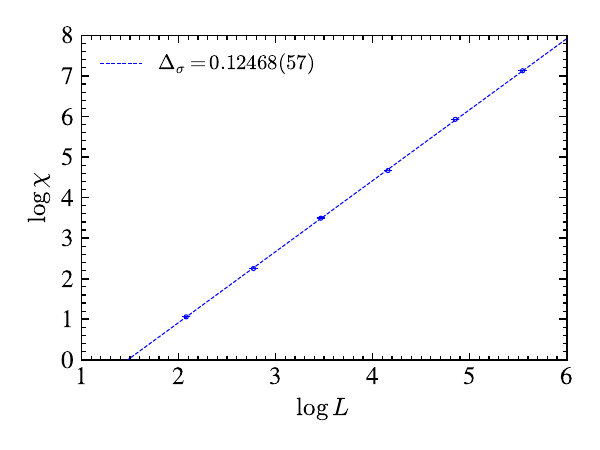}
    \caption{Finite-size dependence of magnetic susceptibility on a lattice with $l_i \propto \{ 4,5,6 \}$.}
    \label{fig:ising_chi_crit}
\end{figure}

We perform a finite-size scaling analysis \cite{Fisher1972ScalingTF,Landau1976FinitesizeBO,Milchev1986FinitesizeSA} to extract the scaling exponent of the spin operator, $\Delta_\sigma$. We measure the magnetic susceptibility $\chi = \langle m^2 \rangle - \langle |m| \rangle^2$ where $m = \sum_i \sigma_i$ is the magnetization. On a finite lattice with characteristic size $L$, the magnetic susceptibility should scale as $\chi(L) \propto L^{2-2 \Delta_{\sigma}}$. Fitting measurements on $L \times L$ triangular lattices with $L = 8$ up to $L = 256$, we find $\Delta_{\sigma} = 0.12468(57)$, in excellent agreement with the exact continuum value $\Delta_{\sigma} = 1/8$. Our fit is shown in Fig. \ref{fig:ising_chi_crit}.

\cleardoublepage

\chapter{Ising model on a 2-sphere}
\label{chapter:sphere}
\thispagestyle{myheadings}

\graphicspath{{4_Sphere/Figures/}}

In this chapter I will use the general relation between simplicial lattice geometry and Ising critical couplings in Eq. (\ref{eq:crit_couplings}) to perform simulations of the critical Ising model on a sequence of simplicial discretizations of a 2-sphere approaching the continuum limit. As I mentioned in Chapter \ref{chapter:simplicial}, there is a finite set simplicial discretizations of a 2-sphere which obey these geometric constraints exactly, which are the simplicial Platonic solids: tetrahedron, octahedron, and icosahedron. Unfortunately, this set is not large enough to perform a reliable continuum extrapolation.

In this chapter, I will first show that for a discretization of $S^2$ with triangles that do not have uniform circumradius and perimeter, the full conformal symmetry of the 2d Ising CFT is \emph{not} recovered in the continuum limit. In Appendix \ref{appendix:iterative_uniform} I describe a method for modifying a discretization of a 2-sphere to minimize non-uniformities in the triangle circumradius and perimeter, and I show that measures of these non-uniformities go to zero in the continuum limit. Performing Monte Carlo simulations of the critical Ising model on these modified lattices, I will show that continuum extrapolations of measurements are in good agreement with the exact solution of the 2d Ising CFT on $S^2$, indicating that the geometric constraints of uniform circumradius and perimeter are indeed necessary to ensure that the model has the desired continuum limit.

\section{Ising model on a 2-sphere}

\begin{figure}
    \centering
    \includegraphics[width=0.3\textwidth]{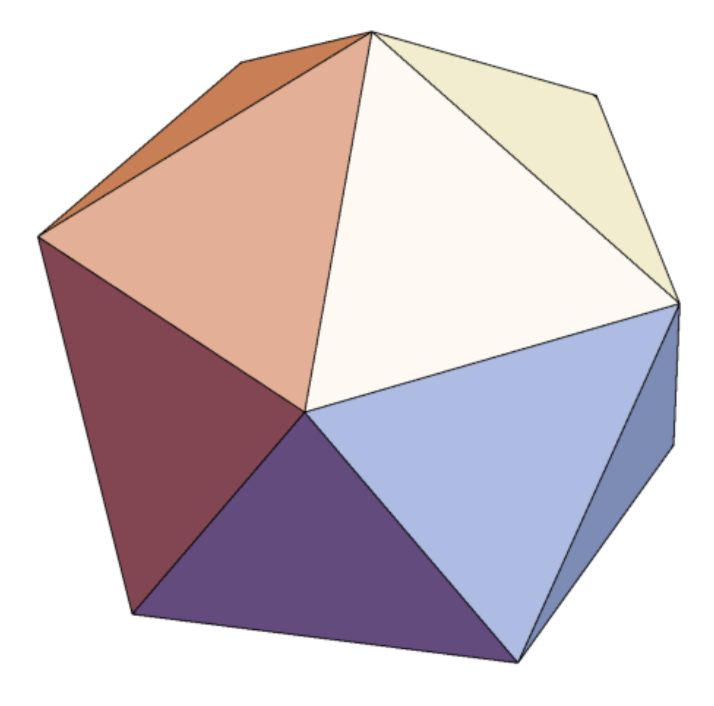}
    \includegraphics[width=0.3\textwidth]{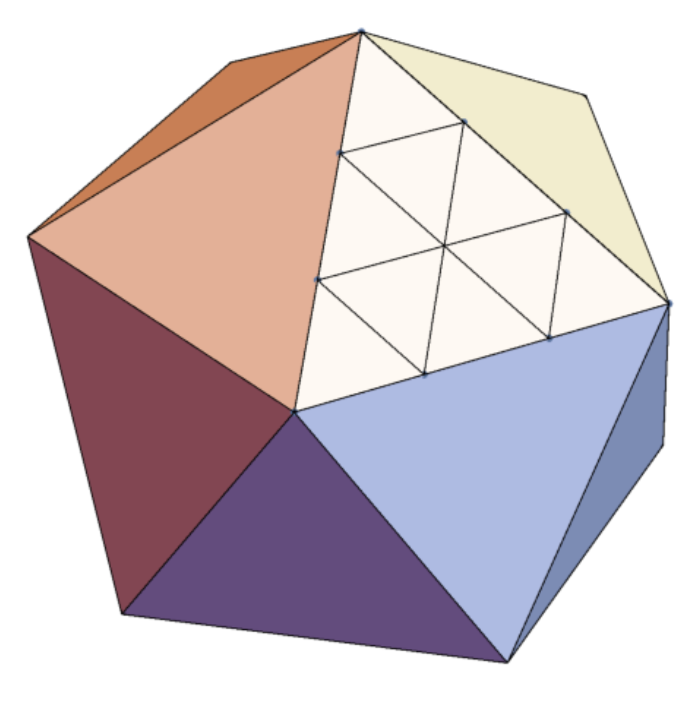}
    \includegraphics[width=0.3\textwidth]{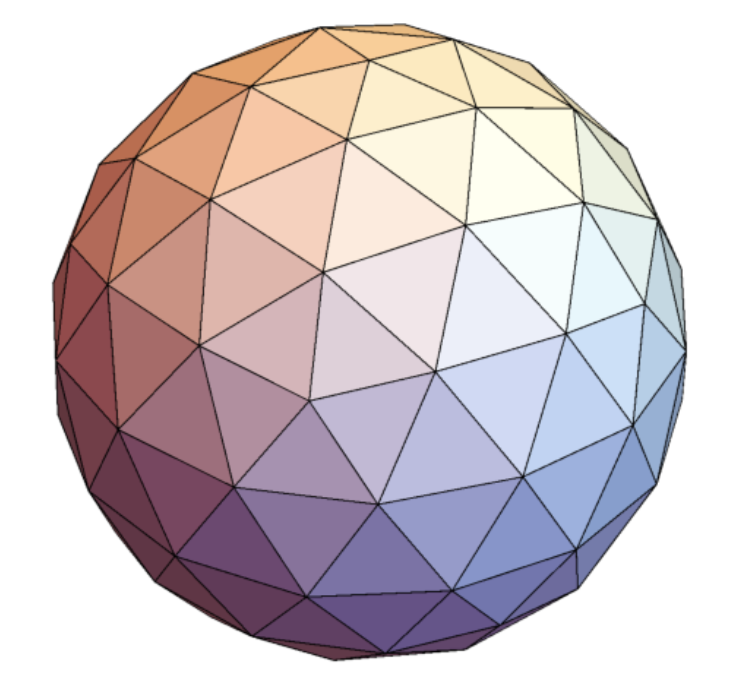}
    \caption{Steps in the basic discretization of $S^2$ using an icosahedral base, shown here for a refinement of 3.}
    \label{fig:ico_refine}
\end{figure}

We define a basic discretization of a 2-sphere constructed by tesselating the faces of one of the simplicial Platonic solids then projecting the resulting vertex coordinates onto a unit sphere, shown in Fig. \ref{fig:ico_refine} for an icosahedron. We use the conventions of Regge calculus to describe the curved geometry of the lattice~\cite{Regge1961GeneralRW}. Specifically, the triangular faces are treated as piecewise flat, with edges acting as ``hinges'' between faces, and all curvature is confined to singularities at the vertices. Although this construction uses unit vectors defined in a 3-dimensional embedding space, the geometry is fully described by the lengths of the lattice edges. These lengths are the fundamental quantities which describe the geometric properties of the manifold, and we note that the sphere and more general simplicial manifolds can be described by Regge calculus without the need for an embedding space.

We use the simplicial Ising action (\ref{eq:tri_action}) with coupling constants $K_{ij}$ for each link calculated using Eq. (\ref{eq:dual_couplings}) and (\ref{eq:crit_couplings}) along with the lattice geometry as defined by the coordinates of the vertices, with dual lattice sites defined at the circumcenters of each triangular face. Because the perimeters of pairs of triangles which share an edge are not always equal we instead define $P_{\triangle}$ in Eq. (\ref{eq:crit_couplings}) as the geometric mean of the perimeters of the two triangles. It is important to note that the dual lattice edge lengths $l_{ij}^*$ are \emph{not} calculated by using the geodesic distance between triangle circumcenters in the 3-dimensional embedding space. Under the conventions of Regge calculus, a metric is only defined on the flat triangular faces of the simplicial complex, and therefore the dual lattice lengths are computed by following straight lines on the flat triangular faces with a ``kink'' when crossing over the corresponding lattice edge.

\begin{figure}
    \centering
    \includegraphics[width=0.35\textwidth]{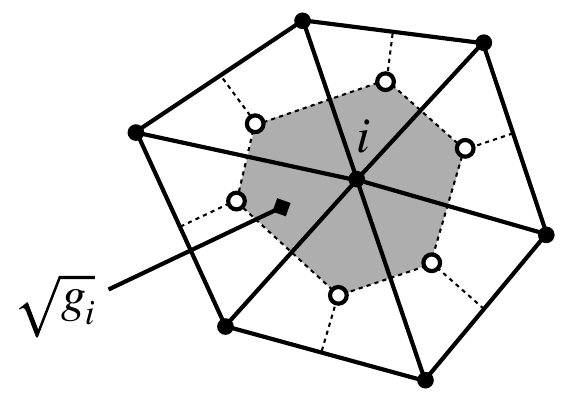}
    \caption{The shaded region is the Voronoi area of lattice site $i$. The open circles are the circumcenters of the adjacent triangles.}
    \label{fig:voronoi}
\end{figure}

We can efficiently measure the spin-spin correlation function on the sphere by projecting onto spherical harmonics
\begin{equation}
    \langle \sigma(\hat n_j) \sigma(\hat n_i) \rangle = \sum_{\ell=0}^{\infty} \sum_{m=-\ell}^{\ell} C_{\ell m} Y^*_{\ell m}(\hat n_i) Y_{\ell m}(\hat n_j)
\end{equation}
where the coefficients
\begin{equation}
    C_{\ell m} = \dfrac{1}{N^2} \left| \sum_i \sqrt{g_i} \sigma_i Y_{\ell m}(\hat n_i) \right|^2
\end{equation}
are the quantities that we measure numerically for each spin configuration generated in our Monte Carlo simulation. Here, $N$ is the number of lattice sites and the quantity $\sqrt{g_i}$ is a discrete integration measure for each site. Following the circumcenter-based conventions of discrete exterior calculus~\cite{Desbrun2005DiscreteEC} it is proportional to the area of the dual simplex associated with the site (i.e. the Voronoi area) as shown in Figure \ref{fig:voronoi}.

We also use the formalism of discrete exterior calculus to define the effective lattice spacing relative to the radius of the sphere. Because we are using a non-uniform lattice, we must adopt a lattice spacing definition which represents a global average over all of the sites. For a lattice discretization of a sphere with $N$ vertices, we simply define the lattice spacing to be equal to the square root of the average Voronoi area, i.e.
\begin{equation}
\label{eq:a_lat_def}
    \dfrac{a^2}{r^2} = \dfrac{1}{N} \sum_i \sqrt{g_i}
\end{equation}
where $r$ is the radius of the sphere and the Voronoi areas are computed using normalized vertex coordinates. This definition of lattice spacing is used to perform all continuum extrapolations on spherical lattices throughout this thesis.

\section{Rotational Symmetry}

The 2-sphere is invariant under the orthogonal group O(3), which has an infinite set of irreducible representations labeled by the familiar $\ell$ index used to describe the spherical harmonics, $Y_{\ell m}(\theta,\phi)$. For the spin-spin correlation function in our lattice model, restoration of rotational symmetry in the continuum limit requires that for each $\ell$, all of the measured $C_{\ell m}$ coefficients for $m \in \{-\ell, ..., \ell\}$ must become degenerate as $a \to 0$. To check this, we define a measurement of rotational symmetry breaking
\begin{equation}
\label{eq:symm_breaking_meas}
    \delta C_{\ell} = 1 - \dfrac{C_{\ell m}^{\textrm{(min)}}}{C_{\ell m'}^{\textrm{(max)}}}
\end{equation}
which captures the maximum deviation between the $C_{\ell m}$ coefficients for a given value of $\ell$.

Because our lattices have been constructed to be symmetric under a discrete subgroup of O(3), some of these coefficients will automatically be degenerate. The full octahedral symmetry group $O_h$ contains a 3-dimensional irrep $T_{1u}$ which maps exactly onto the $\ell=1$ irrep of O(3). Similarly, the full icosahedral symmetry group $I_h$ contains the 3-dimensional and 5-dimensional irreps $T_{1u}$ and $H_g$ which map exactly onto the $\ell=1$ and $\ell=2$ irreps of O(3), respectively. The corresponding coefficients of the 2-point function on lattices with octahedral or icosahedral symmetry do not break rotational symmetry (up to statistical errors), and therefore they provide a good indication for what the symmetry-breaking measurement should look like for an unbroken irrep.

\begin{figure}
    \centering
    \includegraphics{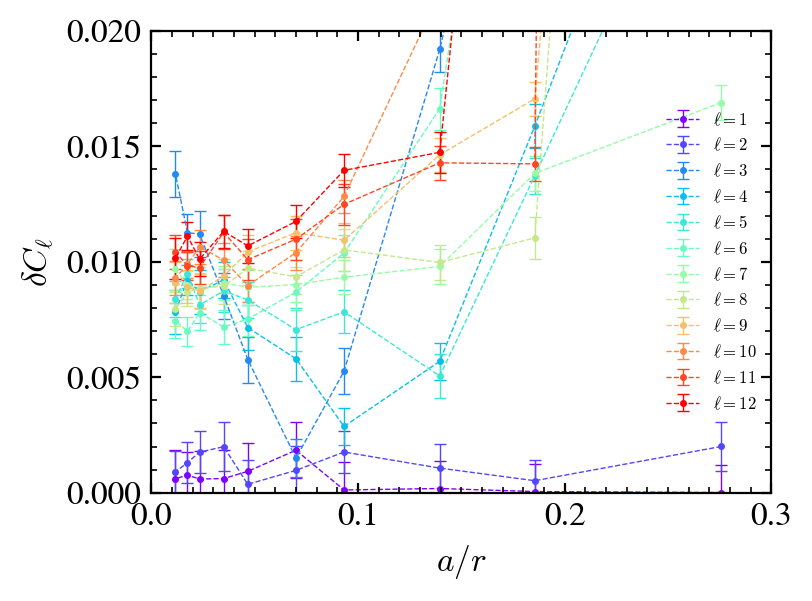}
    \caption{Rotational symmetry breaking of the spin-spin correlation function using the basic icosahedral discretization of $S^2$ up to a refinement of 96.}
    \label{fig:q5_rot_break_naive}
\end{figure}

In Fig. \ref{fig:q5_rot_break_naive} we show the rotational symmetry breaking measurement as a function of the lattice spacing using the basic discretization of the icosahedron. The measurement is clearly approaching a nonzero value in the continuum limit for $\ell \geq 2$, which indicates that this construction does not restore rotational symmetry.
Because the triangular faces of the basic discretization of the 2-sphere fail to satisfy the constraints of equal circumradius and equal perimeter, it is perhaps unsurprising that the higher $\ell$ coefficients of the spin-spin correlation function do not converge to zero in the continuum limit. One might have hoped that because the triangular faces in the basic discretization have \emph{locally} uniform circumradius and perimeter (i.e. deviations in the circumradius and perimeter of neighboring triangles go to zero in the continuum limit), this might have been sufficient to satisfy the geometric constraints required by our derivation in Chapter \ref{chapter:simplicial}. However, it is clear from our measurements here that variations in triangle geometry over long distances result in a lattice theory which fails to fully restore continuum symmetries. We therefore conclude that the critical couplings derived in Chapter \ref{chapter:simplicial} do not result in a lattice theory with a well-defined continuum limit if these geometrical constraints are only satisfied locally.

\begin{figure}
    \centering
    \includegraphics{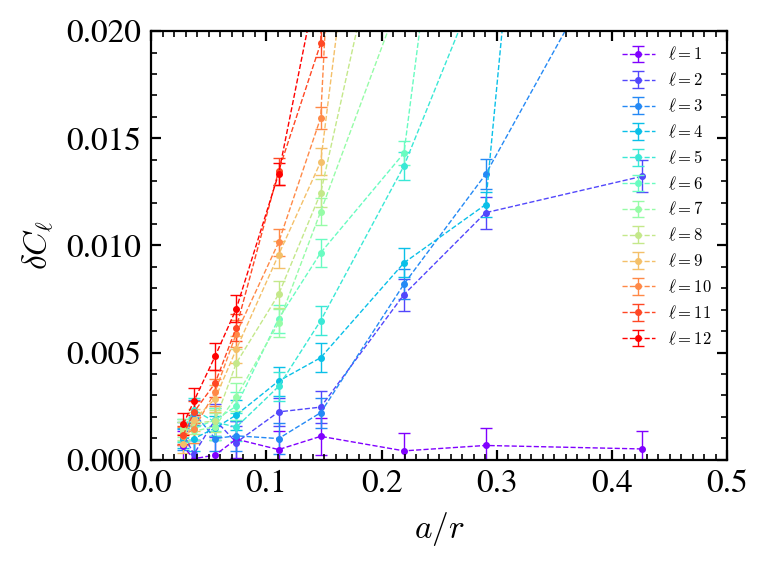}
    \includegraphics{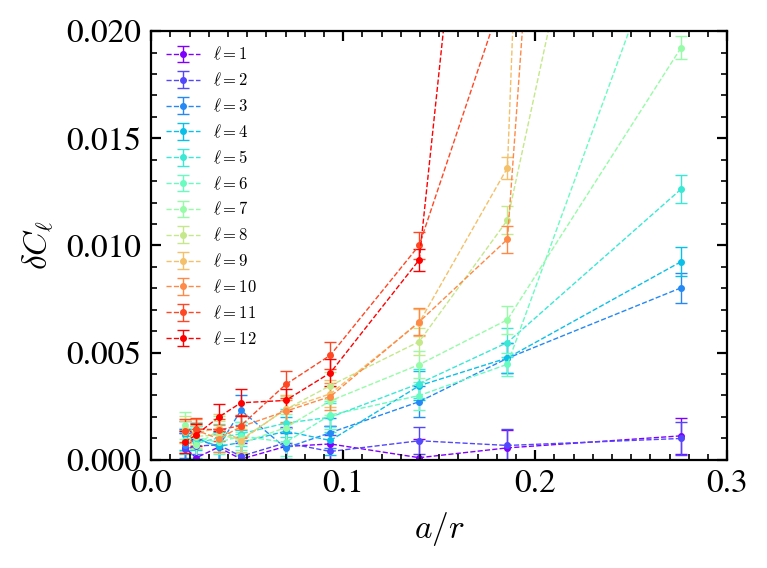}
    \caption{Rotational symmetry breaking of the spin-spin two-point function as a function of lattice spacing using the modified discretization of the sphere using an octahedral (top) and icosahedral (bottom) base lattice up to a refinement of 64.}
    \label{fig:rot_break_mod}
\end{figure}

To remedy this issue, we repeat the Monte Carlo simulations of the previous section, but this time we use the modified lattice construction described in Appendix \ref{appendix:iterative_uniform}, which explicitly minimizes non-uniformities in the circumradius and perimeter of triangular faces so that \emph{global} variations in these quantities go to zero in the continuum limit. We again plot the symmetry breaking measurement $\delta C_{\ell}$ as a function of lattice spacing in Fig. \ref{fig:rot_break_mod}. 
For both the octahedral and icosahedral discretizations, the rotational symmetry-breaking measurement goes to zero for all measured irreps within statistical uncertainty. This result indicates that global uniformity in the triangle geometry is necessary to generate a sequence of lattices for which this model restores rotational symmetry in the continuum limit. It is especially promising that this procedure works even for the octahedral lattice, which requires much larger variations in triangle shape in order to fully tesselate the sphere, which can be seen explicitly in Fig. \ref{fig:lattice_uniform}.

\begin{figure}
    \centering
    \includegraphics{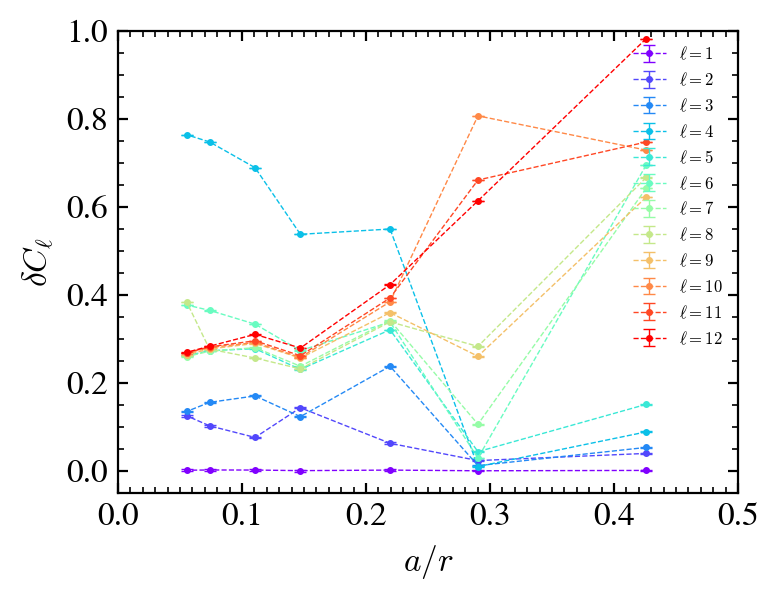}
    \caption{Severe breaking of rotational symmetry for an octahedral base lattice after applying the optimization procedure of \cite{Tegmark1996AnIM}.}
    \label{fig:q4_rot_break_tegmark}
\end{figure}

It is worth noting that although the triangle circumradius and perimeter are the quantities that appeared in our derivation of the Ising critical couplings, in general there are many other ways to define the measure of uniformity for a simplicial lattice. As an example, we tested our simulation of the Ising model on a discretized sphere using the optimization procedure described in \cite{Tegmark1996AnIM}, which adjusts the positions of the lattice sites so that the Voronoi areas of all sites become approximately equal. However, just as we saw when using the basic discretization, this construction fails to restore rotational symmetry in the continuum limit. The rotational symmetry breaking is especially strong using a octahedral base lattice, as shown in Fig. \ref{fig:q4_rot_break_tegmark}. We also tested several other methods for adjusting the vertices with similar results~\cite{Xu2006DiscreteLO,Ahrens2009RotationallyIQ,Iga2014ImprovedSA,Fornberg2014OnSH}. We therefore conclude that the definition of uniformity based on the circumradius and perimeter is indeed necessary to ensure a valid continuum limit for the critical Ising model.

\section{Agreement with the Ising CFT}

\begin{figure}
    \centering
    \includegraphics{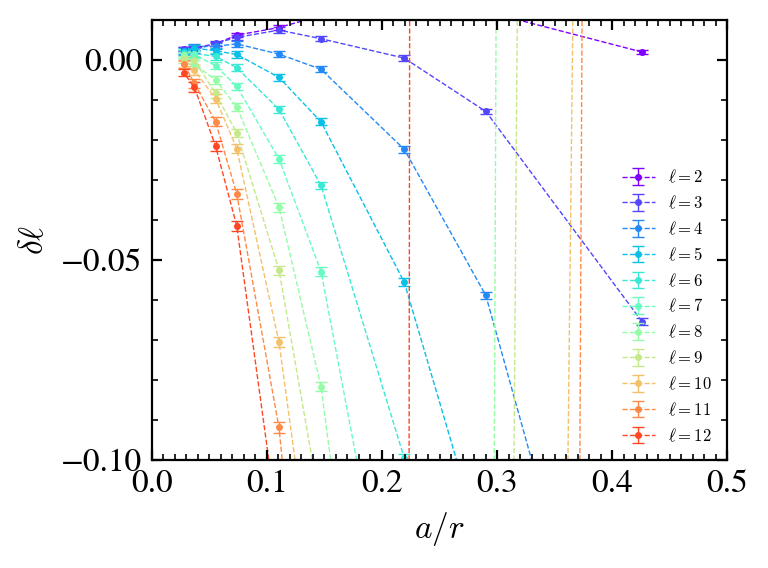}
    \includegraphics{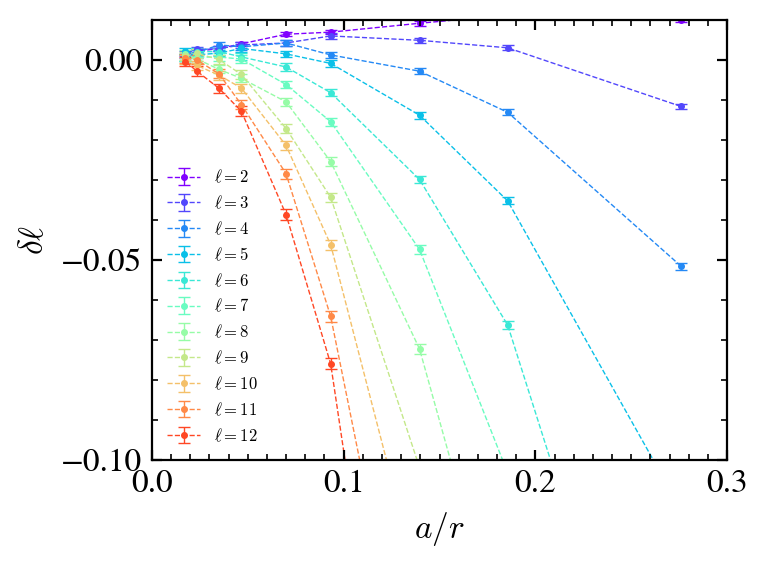}
    \caption{Conformal symmetry breaking measurement for the spin-spin two-point function as a function of lattice spacing using the modified discretization of the sphere using an octahedral (top) and icosahedral (bottom) base lattice up to a refinement of 64.}
    \label{fig:cft_break_mod}
\end{figure}

Now that we have confirmed that the spin-spin correlation function in our lattice theory becomes rotationally symmetric in the continuum limit, we would like to check that it also agrees with the exactly known analytical result for the 2d Ising CFT on a 2-sphere as derived in Appendix \ref{appendix:embedding_space}:
\begin{equation}
    \langle \sigma_i \sigma_j \rangle \propto \dfrac{1}{(1 - \hat n_i \cdot \hat n_j)^{\Delta_{\sigma}}} = \sum_{\ell} \dfrac{2 \ell + 1}{2} F_{\ell} P_{\ell} (\hat n_i \cdot \hat n_j)
\end{equation}
where $\Delta_{\sigma} = 1/8$ and we have expanded in a series of Legendre polynomials which have coefficients
\begin{equation}
\label{eq:2pt_coeff}
    \dfrac{F_{\ell}}{F_0} = \dfrac{\Gamma(\Delta_{\sigma} + \ell) \Gamma(2 - \Delta_{\sigma})}{\Gamma(\Delta_{\sigma}) \Gamma(2 - \Delta_{\sigma} + \ell)}
\end{equation}
which we can measure directly from lattice configurations generated by our Monte Carlo simulations. Eliminating $\Delta_{\sigma}$ and solving for $\ell$ we find
\begin{equation}
    \ell = \dfrac{(F_0 - F_1)(F_{\ell-1} + F_{\ell})}{(F_0 + F_1)(F_{\ell-1} - F_{\ell})}\;.
\end{equation}
To check that the lattice simulation agrees with the analytic result, we first calculate a conformal symmetry breaking measurement
\begin{equation}
    \delta \ell = 1 - \dfrac{(F_0 - F_1)(F_{\ell-1} + F_{\ell})}{\ell (F_0 + F_1)(F_{\ell-1} - F_{\ell})}
\end{equation}
which should go to zero in the continuum limit. This quantity is plotted as a function of lattice spacing in Fig. \ref{fig:cft_break_mod}, showing the expected behavior for both the octahedral and icosahedral lattice for all irreps of O(3) up to $\ell = 12$.

\begin{figure}
    \centering
    \includegraphics{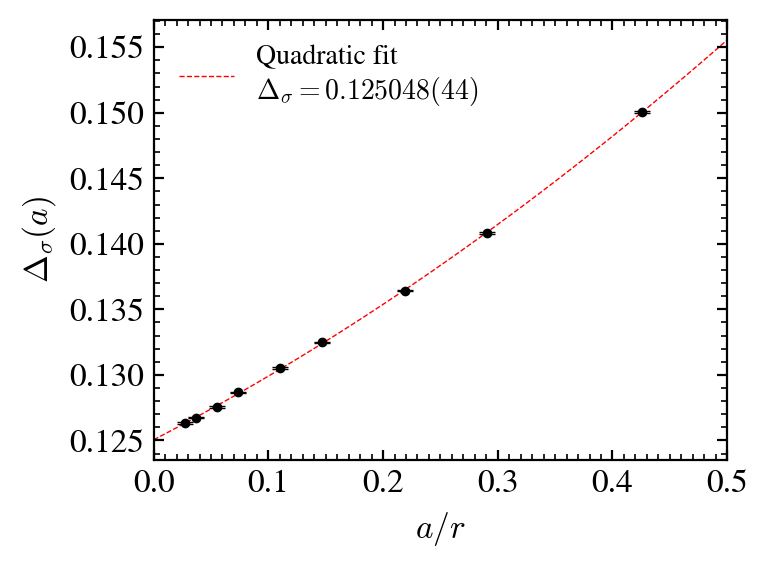}
    \includegraphics{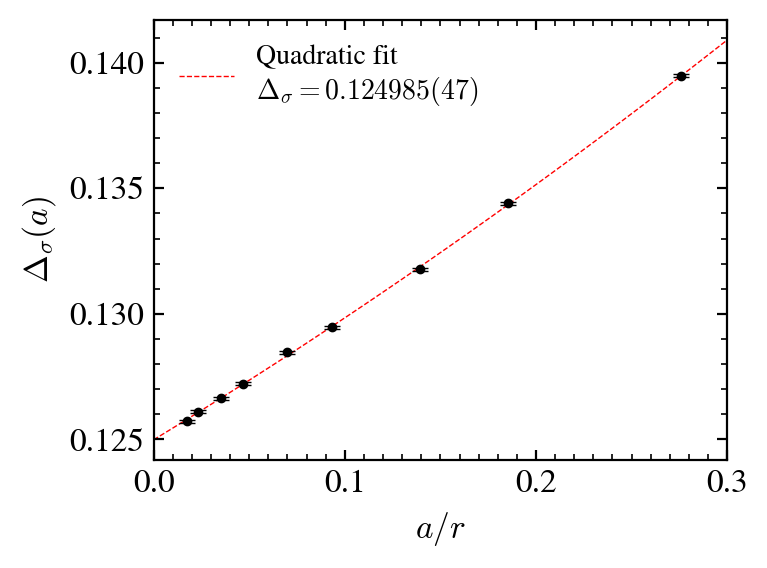}
    \caption{Continuum extrapolation of the scaling exponent $\Delta_{\sigma}$ for the modified octahedral (top) and icosahedral (bottom) lattice.}
    \label{fig:sphere_delta_sigma}
\end{figure}

Finally, we measure the scaling exponent of the $\sigma$ operator on the lattice via
\begin{equation}
    \Delta_{\sigma} = \dfrac{2 F_1}{F_1 + F_0} 
\end{equation}
which is plotted as a function of lattice spacing in Fig. \ref{fig:sphere_delta_sigma}. Extrapolating to the continuum limit by fitting to a quadratic polynomial in $a$, we obtain the results $\Delta_{\sigma} = 0.125048(44)$ with $\chi^2/\text{dof}=1.8$ for the octahedral lattice and $\Delta_{\sigma} = 0.124985(47)$ with $\chi^2/\text{dof}=1.8$ for the icosahedral lattice, both in excellent agreement with the exact value of $1/8$ and with a relative uncertainty of about $0.03\%$.

Our results indicate that by using the modified discretization of $S^2$, our lattice simulation accurately captures the properties of the lowest $\mathbb{Z}_2$-odd operator $\sigma$. The 2d Ising CFT of course contains an infinite set of operators related by the Virasoro algebra. Measurement of additional operators using our lattice model is a subject for future study outside the scope of this thesis, however based on the high level of accuracy of our results and the delicate nature of a properly-tuned conformal field theory, we believe it is likely that our lattice formulation is sufficient to capture the properties of all of the operators in the 2d Ising CFT in the continuum limit, given sufficient computational resources.

\begin{figure}
    \centering
    \includegraphics{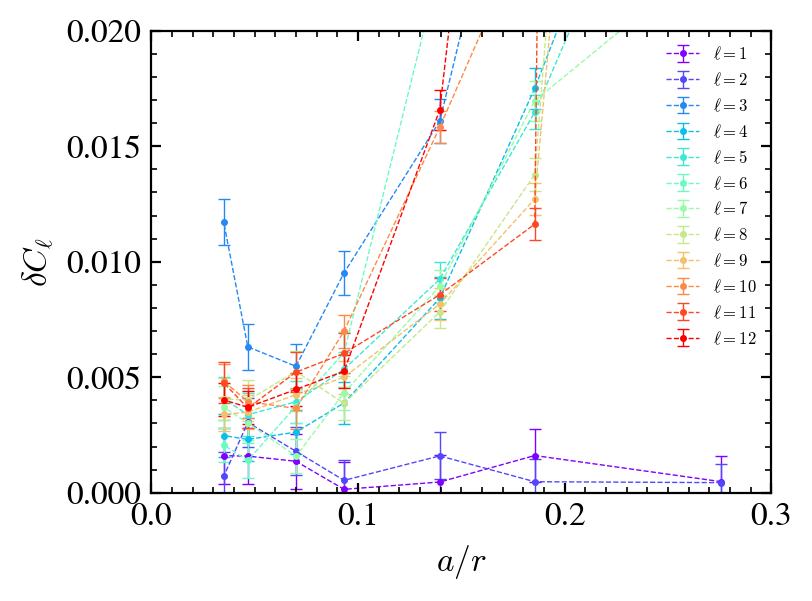}
    \caption{Breaking of rotational symmetry in critical $\phi^4$ theory with local perturbative counterterms using the basic icosahedral discretization of $S^2$.}
    \label{fig:phi4_break}
\end{figure}

It is worth noting that the restoration of rotational symmetries in a lattice model of a sphere is a challenging problem which has been encountered in previous works. In \cite{Brower2012LatticeRQ} attempts were made to simulate the 3d Ising CFT on the manifold $\mathbb{R} \times S^2$ of radial quantization using an icosahedral discretization of the sphere and uniform couplings. Although this construction was shown to have a well-defined critical point, rotational symmetry was not restored in the continuum limit. In \cite{Brower2018LatticeF,Brower2020RadialLQ}, simulations of critical $\phi^4$ theory were performed using a generalization of the finite element method. Again, it was shown that a well-defined critical point exists after adding a perturbative mass counterterm to account for the non-uniform UV divergence inherent in the lattice. However, as shown in Fig. \ref{fig:phi4_break}, higher precision measurements of the scalar 2-point function reveal a slight breaking of rotational symmetry in the continuum limit, very similar to the same results presented in Fig. \ref{fig:q5_rot_break_naive} for the Ising model on the basic discretization of the sphere. It therefore seems likely that some additional geometric constraints may be necessary in order to restore rotational symmetry in these theories as well. It's possible that the same constraints used here (uniform triangle circumradius and perimeter) may also work for other theories, but although we have begun to study this possibility it is unclear if this is the case.

Another approach taken recently is to use the so-called ``fuzzy sphere'' which projects the states of the theory in the basis of spherical harmonics instead of using a traditional spatial lattice~\cite{Zhu2022UncoveringCS,Hu2023OperatorPE}. In this case, the lattice cutoff is replaced by a maximum spherical harmonic quantum number, which is a more manifest way to ensure that rotational symmetry is restored in the continuum limit. This method shows promising results for the lowest operators of the 3d Ising CFT, though it is unclear how difficult it would be to generalize to other theories.
\cleardoublepage

\chapter{Conclusion}
\label{chapter:conclustion}
\thispagestyle{myheadings}

\graphicspath{{5_Conclusion/Figures/}}

In Chapter \ref{chapter:intro}, I described how lattice simulations of conformal field theories can be greatly improved by using discretizations of the radial quantization ``cylinder'' $\mathbb{R} \times S^{d-1}$ and the sphere $S^d$, when compared to traditional simulations which use a $d$-dimensional torus. The challenge of using these spherical manifolds is that it is difficult to define a discrete action on a lattice which restores the full rotational symmetry of the sphere in the continuum limit. To overcome this difficulty for the 2d Ising model, I derived a relation between the geometry of a simplicial discretization of a 2-sphere and the critical coupling constants, and I showed that the 2d Ising CFT can indeed be recovered in the continuum limit. In this chapter I will discuss how similar methods may exist for other theories and in higher dimensions.


Given that the research in this thesis only applies to the exactly soluble 2d Ising CFT, a good deal of skepticism is anticipated in regards to whether or not similar methods exist for other theories. One might ask whether a non-uniform simplicial lattice formulation which restores continuum symmetries as the lattice spacing goes to zero even exists. While I don't claim to be able to answer this question for any specific theories, I can state two non-trivial requirements that we have learned from the 2d Ising model.

First, it was necessary to obtain a map which gives a one-to-one correspondence between the lattice geometry and the parameters of the simplicial lattice action. As I showed in Chapter \ref{chapter:affine}, this required a 2-parameter map to obtain the proper coupling constants. This corresponds with the number of extra parameters for a general affine transformation (besides translation, rotation, and scaling). Generalizing to $d$-dimensions, a general affine transformation requires a total of $d(d+1)$ parameters. Translations and rotations account for $d$ parameters and $d(d-1)/2$ parameters, respectively, while scaling requires a single parameter. The remaining $(d+2)(d-1)/2$ parameters are in one-to-one correspondence with the geometries of all possible $d$-simplices. For example, a 3-dimensional theory on a non-uniform simplicial lattice would require the proper determination of a 5-parameter map between the lattice geometry and the parameters in the lattice action.

Even if no such map can be derived analytically as we have done here, it may be possible to compute such a map numerically. In fact, this is exactly the approach used to compute the so-called ``Karsch coefficients''~\cite{Karsch1982SUNGT} used in finite temperature lattice QCD which describe the non-perturbative relationship between the coupling constants and the lattice spacing on an anisotropic hypercubic lattice. Indeed, the 1-parameter map defined by Karsch coefficients is a subset of the 9-parameter map that would be required to simulate QCD on an arbitrary 4d simplicial lattice.

The second requirement that we discovered was the necessity for the lattice geometry to be globally uniform. This accounts for the last parameter in an arbitrary affine transformation which describes the overall scale of each simplex. In the case of the 2d Ising model, we found that all of the triangular faces must have the same circumradius and perimeter. Although it is not possible to satisfy this constraint exactly on an arbitrary discretization of a 2-sphere, in Chapter \ref{chapter:sphere} I showed that it is sufficient to ensure that deviations in the uniformity of the triangles go to zero in the continuum limit.

Although the lattice Ising model is not a field theory, from the perspective of quantum field theory it is tempting to recast these geometrical constraints as requirements about UV divergences. It is a well-known property of any interacting quantum field theory that the UV behavior must be regulated in order to extract information about the physical properties of the theory. Most regulators (i.e. a momentum cutoff, Pauli-Villars regularization, dimensional regularization, and the traditional square lattice) respect the continuum symmetries of the theory (or at least a discrete subgroup of the continuum symmetries). However, in the case of a non-uniform lattice, it's unclear what effect the non-uniformity of the regulator will have on the UV properties of the theory. With this in mind, one interpretation of the results of this thesis is that the geometrical constraints required to restore continuum symmetries in the critical Ising model are somehow related to the requirement that a lattice regulator must be uniform in order to properly describe the UV divergences of a quantum field theory. An interesting topic for future study is therefore to determine the general constraints that a non-uniform lattice must satisfy in order for a well-defined quantum field theory to emerge in the continuum limit.

There are several obvious future extensions of the work presented here. First, it would be interesting to find a simplicial generalization of the 3d Ising model. Unlike the 2d Ising model, which is ``self-dual'', it is unlikely that duality relations can be used to derive analytic relations between the lattice geometry and the 3d Ising coupling constants. However, it may be possible to determine these relations numerically, and attempts to develop a procedure for this are underway. A successful 3d simplicial lattice model for the critical 3d Ising model would allow for simulations on the manifolds $S^3$ and $\mathbb{R} \times S^2$ in order to study the 3d Ising CFT. Such studies would be an excellent test of the conformal bootstrap method, which places strong constraints on the properties of the 3d Ising CFT, albeit via rather indirect methods related to the exploitation of conformal symmetry~\cite{ElShowk2012SolvingT3,Kos2014BootstrappingMC}.

Another interesting extension would be to develop methods for performing gauge theory simulations on simplicial lattices. As I mentioned earlier, anisotropic lattices used in finite-temperature QCD require unequal lattice spacings to be determined numerically via Monte Carlo simulations. Extension of these methods to general simplicial manifolds would allow for a wide range of new applications. For example, a 3-dimensional U(1) gauge theory coupled to matter fields, known as 3d QED, is known to have a critical point described by a CFT~\cite{Appelquist1988CriticalBI} which is not well-understood. There also exist conformal and near-conformal QCD-like theories which can possibly describe a composite Higgs boson~\cite{Appelquist2020NearconformalDI,Appelquist2021GoldstoneBS}. Simplicial lattice methods for these and other theories would mitigate the finite-volume obstructions inherent in traditional lattice simulations of conformal theories, providing more direct access to the physical properties of these interesting continuum theories.

While it's true that generalizations of the methods developed in this thesis will be theoretically challenging and computationally expensive to develop, based on the high accuracy of the numerically results that I have presented here for the 2d Ising model, it appears that it would be worthwhile to pursue such methods further.

\cleardoublepage

\begin{appendices}
\chapter{Some properties of conformal field theory}
\label{appendix:embedding_space}
\thispagestyle{myheadings}

\graphicspath{{Appendix/Figures/}}

In this appendix I will describe some general properties of conformal field theory that are referenced elsewhere in this thesis. For a much more thorough reference on conformal field theory, see \cite{Francesco1999ConformalFT} and \cite{Rychkov:2016iqz} which contain all of the information presented here. Also note that throughout this thesis I consider conformal field theory in Euclidean space rather than Minkowski space.

\section{Conformal Algebra}

Conformal field theories are invariant under conformal transformations. These transformations form the conformal group, which itself contains the translations and rotations of the Euclidean group as a proper subgroup, as well as additional transformations known as dilatations and special conformal transformations. The generators of conformal transformations are
\begin{equation}
\begin{split}
    P_{i} &= -i \partial_{\mu} \\
    D &= -i x^{i} \partial_{i} \\
    L_{ij} &= i (x_{i} \partial_{j} - x_{j} \partial_{i}) \\
    K_{i} &= -i ( 2 x_{i} x^{j} \partial_{j} - x^2 \partial_{i} )
\end{split}
\end{equation}
for translations, dilatations, rotations, and special conformal transformations, respectively. The conformal algebra is defined by the commutation relations
\begin{equation}
\begin{split}
    [D,P_i] &= i P_i \\
    [D,K_i] &= -i K_i \\
    [K_i,P_j] &= 2i (\delta_{ij} D - L_{ij}) \\
    [K_k,L_{ij}] &= i (\delta_{ik} K_j - \delta_{jk} K_i) \\
    [P_k,L_{ij}] &= i( \delta_{ik} P_j - \delta_{jk} P_i) \\
    [L_{ij},L_{k\ell}] &= i( \delta_{jk} L_{i \ell} + \delta_{i \ell} L_{j k} - \delta_{ik} L_{j \ell} - \delta_{j \ell} L_{ik} )
\end{split}
\end{equation}
with all other commutators being zero. The special conformal transformations are related to the translations via
\begin{equation}
    K_i x = I P_i I x
\end{equation}
where $I$ is the inversion operator:
\begin{equation}
    I x^i = \dfrac{x^i}{x^2}
\end{equation}

\section{Embedding Space Formalism}

In order to understand the conformal group in the context of CFTs defined on various curved manifolds, we will introduce the embedding space formalism. We combine the conformal generators into an antisymmetric matrix $J_{AB}$ with $A,B \in \{ -1, 0, 1, \dots, d\}$
\begin{equation}
\begin{split}
    J_{ij} &= L_{ij} \\
    J_{0,i} &= \dfrac{1}{2}(P_i + K_i) \\
    J_{-1,i} &= \dfrac{1}{2}(P_i - K_i) \\
    J_{-1,0} &= D
\end{split}
\end{equation}
Together with Minkowski metric $\eta_{AB} = \text{diag}(-1,1,\dots,1)$, the generators $J_{AB}$ obey the commutation relations of the Lie group $SO(d+1,1)$:
\begin{equation}
    [J_{AB},J_{CD}] = i( \eta_{AD} J_{BC} + \eta_{BC} J_{AD} - \eta_{AC} J_{BD} - \eta_{BD} J_{AC} ) \;.
\end{equation}
Thus, we have the important fact that the conformal group in $d$-dimensions is isomorphic to the Lorentz group $SO(d+1,1)$.

In order to make sense of this isomorphism in terms of a $d$-dimensional conformal field theory, we can map the points in the $d$-dimensional space onto the light rays of $\mathbb{R}^{d + 1,1}$ by choosing any continuous cross section of the light cone. Under a Lorentz transformation, light rays transform into other light rays. Therefore a conformal transformation can be defined by the equivalent Lorentz transformation on the embedding space, followed by a projection back to the original cross section. This is known as the \emph{projective light-cone} (or null-cone) formalism. This can be further understood by noting that the set of cross sections are manifolds which transform into one another under Weyl transformations.

\begin{figure}
    \centering
    \includegraphics[width=0.65\textwidth]{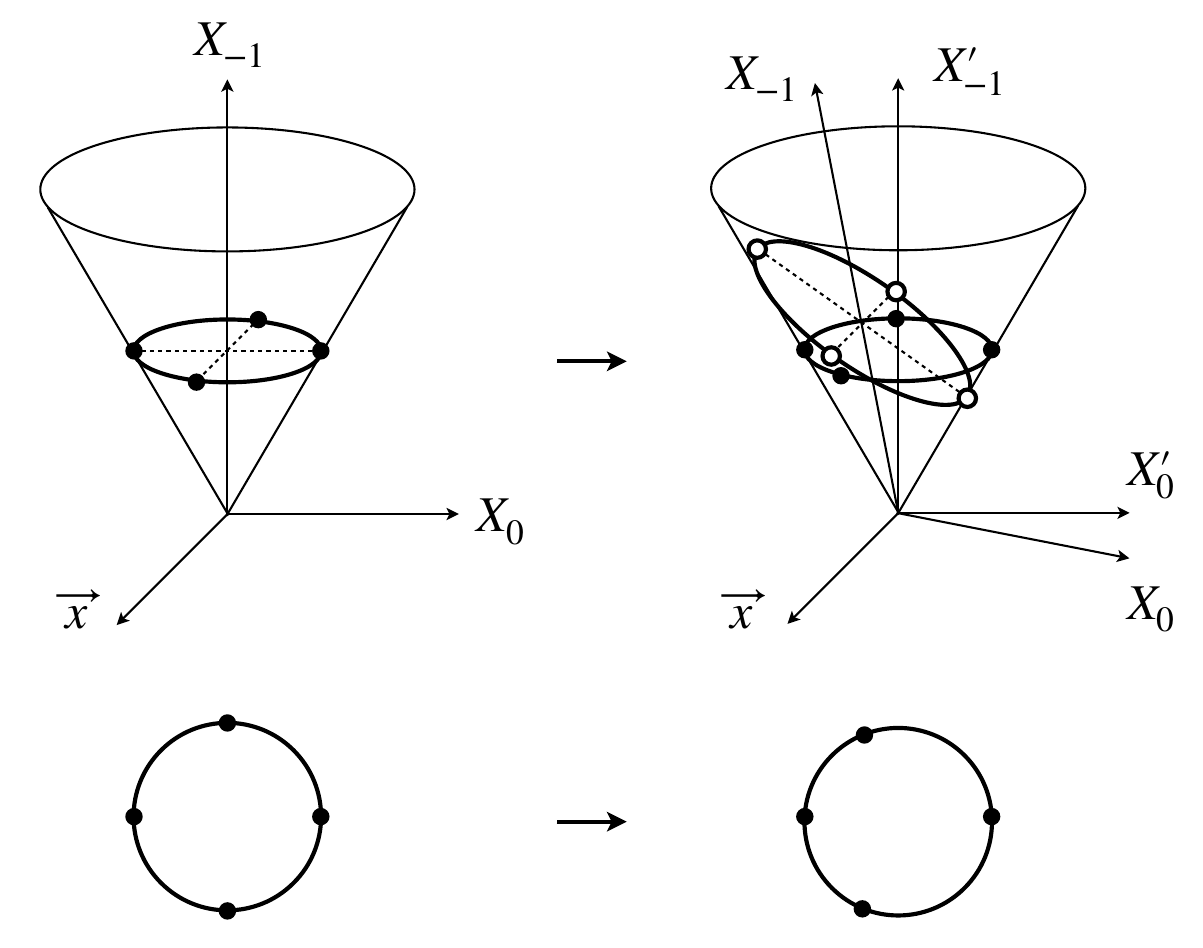}
    \caption{Effect of the dilatation operator on a spherical cross-section of the light cone. The $d$-dimensional sub-manifold has been compactified into a single axis for visual clarity.}
    \label{fig:cft_lorentz_transform}
\end{figure}

As an example, we consider the effect of the dilatation operator on a CFT defined on a spherical cross-section of the light cone, pictured in Fig. \ref{fig:cft_lorentz_transform}. The corresponding Lorentz transformation in the $(d+1,1)$-dimensional Minkowski space is a boost in the $X^0$ direction. The boost transforms the sphere on the left into an ellipsoid in the new coordinate system (open circles). After using light rays to project back onto the spherical cross section (filled circles), we see that the overall effect of the boost is that the points on the sphere have all shifted to the left. This is in contrast to the effect of the dilatation operator on the infinite plane, which simply produces a uniform rescaling of the entire manifold.

\section{Invariant distance}

The metric of the embedding space defines an invariant distance between two points $X$ and $Y$ in the embedding space:
\begin{equation}
    \eta_{AB} X^A Y^B = X^A Y_A\;.
\end{equation}
The light cone is defined to be the set of null-vectors, i.e. all points $X^A$ such that $X_A X^A = 0$.

There are an infinite number of possible cross-sections on which we can define a conformal field theory, so we will describe a few important examples here. To recover the flat-space manifold $\mathbb{R}^d$, we choose as our cross-section the set of points
\begin{equation}
    X = \left( \dfrac{1 + |\vec x| ^2}{2}, \dfrac{1 - |\vec x| ^2}{2}, \vec x \right)
\end{equation}
where $\vec x \in \mathbb{R}^d$. Up to a constant of proportionality, the invariant distance between two points is then
\begin{equation}
    X^A Y_A \propto |\vec x - \vec y|^2 \;.
\end{equation}

Another useful cross-section is the sphere $S^d$, which is defined as a cross-section of the light cone by the points
\begin{equation}
    X = (1, \hat n)
\end{equation}
where the $\hat n$ is a $(d+1)$-dimensional unit vector in the $\mathbb{R}^{d+1}$ sub-manifold of the embedding space. The invariant distance for points on the sphere is
\begin{equation}
    X^A Y_A \propto 1 - \hat n_x \cdot \hat n_x \;.
\end{equation}

One last cross-section that we will consider here is the ``cylinder'' of radial quantization, which has the cross section of a $(d-1)$-sphere. It is defined by the points
\begin{equation}
    X = (\cosh t, \sinh t, \hat n)
\end{equation}
where $(t, \hat n) \in \mathbb{R} \times S^{d-1}$. This manifold is particularly useful for the study of conformal field theories because the dilatation operator generates translations in the radial coordinate $\tau$ while the rotation operator generates the rotations of the sphere. This allows the physical states of the theory to be easily decomposed into states which are simultaneous eigenstates of both the dilatation operator and angular momentum. The invariant distance on this manifold is
\begin{equation}
    X^A Y_A \propto \cosh (t_x - t_y) - \hat n_x \cdot \hat n_x \;.
\end{equation}

\section{Correlation functions}

Conformal symmetry constrains correlation functions for operators in a conformal field theory to be of the form
\begin{equation}
    \langle \mathcal{O}(X) \mathcal{O}(Y) \rangle \propto \dfrac{1}{(X^A Y_A)^{\Delta_{\mathcal{O}}}}
\end{equation}
where $\Delta_{\mathcal{O}}$ is the scaling exponent of the operator $\mathcal{O}$. Using the definitions of $X^A Y_A$ for the different choices of cross-section we therefore know the form of a conformal correlator on these different manifolds. On the infinite plane we simply have
\begin{equation}
    \langle \mathcal{O}(\vec x) \mathcal{O}(\vec y) \rangle_{\mathbb{R}^d} \propto \dfrac{1}{|\vec x - \vec y|^{2 \Delta_{\mathcal{O}}}}
\end{equation}
where the normalization is typically chosen such that the constant of proportionality is 1. Note that on the lattice, the normalization must be determined using lattice measurements. On the sphere we have
\begin{equation}
    \langle \mathcal{O}(\hat n_x) \mathcal{O}(\hat n_y) \rangle_{S^d} \propto \dfrac{1}{(1 - \hat n_x \cdot \hat n_y)^{\Delta_{\mathcal{O}}}}
\end{equation}
and for the cylinder of radial quantization we have
\begin{equation}
    \langle \mathcal{O}(t_x, \hat n_x) \mathcal{O}(t_y, \hat n_y) \rangle_{\mathbb{R} \times S^{d-1}} \propto \dfrac{1}{(\cosh (t_x - t_y) - \hat n_x \cdot \hat n_y)^{\Delta_{\mathcal{O}}}}\;.
\end{equation}

These forms allow us to efficiently extract the scaling exponents from measurements of lattice simulations. For example, a correlator for a 2-dimensional CFT in radial quantization can be expanded as a Fourier cosine series
\begin{equation}
    \langle \mathcal{O}(t_x, \hat n_y) \mathcal{O}(t_x, \hat n_y) \rangle_{S^1 \times \mathbb{R}} = \dfrac{1}{2} C_{0}(t) + \sum_{m=1}^{\infty} C_{m}(t) \cos (m \theta)
\end{equation}
where $t = |t_x - t_y|$ and $\cos \theta = \hat n_x \cdot \hat n_y$. Using $C_0(0)$ to fix the normalization, the coefficients are
\begin{equation}
\label{eq:rad_2pt_full}
    \dfrac{C_m(t)}{C_0(0)} = \dfrac{\Gamma(1-\Delta_{\mathcal{O}})^2}{\Gamma(1-2\Delta_{\mathcal{O}})} \sum_{n=0}^{\infty} \dfrac{(\Delta_{\mathcal{O}})_n (\Delta_{\mathcal{O}})_{n+m} }{n! (n+m)! }   e^{-(\Delta_{\mathcal{O}} + 2n + m)t}
\end{equation}
where $(\Delta_{\mathcal{O}})_n = \Gamma(\Delta_{\mathcal{O}} + n) / \Gamma(\Delta_{\mathcal{O}})$ is the Pochhammer symbol. The scaling exponent $\Delta_{\mathcal{O}}$ can be extracted from lattice measurements of these coefficients by fitting the large $t$ data to an exponential. Or, for an equal-time correlator with $t=0$ this becomes
\begin{equation}
    \dfrac{C_m(0)}{C_0(0)} = \dfrac{(\Delta_{\mathcal{O}})_m}{(1 - \Delta_{\mathcal{O}})_{m}}
\end{equation}
which can also be used to extract $\Delta_{\mathcal{O}}$. The coefficients $C_m(t)$ can be treated as bulk measurements on the lattice, and can therefore be measured very accurately compared to direct measurement of the 2-point function. An expansion of a conformal correlator on $S^2$ is given in chapter \ref{chapter:sphere} in terms of Legendre polynomials, and similar expansions can be derived for other manifolds.

\chapter{Iterative method for generating uniform spherical meshes}
\label{appendix:iterative_uniform}
\thispagestyle{myheadings}

In Chapter \ref{chapter:sphere}, I showed that rotational symmetry of the spin-spin correlation function was broken in the continuum limit for simulations of the critical Ising model on the basic discretization of a sphere. The failure to restore rotational symmetry in the continuum limit can be traced back to the fact that our derivation of the critical couplings on a simplicial lattice required all of the triangles to have equal circumradius and perimeter, which is not the case for the basic discretization of the sphere. In this appendix I will describe an iterative method for modifying the basic discretization of $S^2$ to make the triangles more uniform.

\section{Measure of non-uniformity}

The only simplicial discretizations of a 2-sphere which can be constructed from uniform triangles are the platonic solids: tetrahedron, octahedron, and icosahedron\footnote{The other two platonic solids, the cube and dodecahedron, are not included here because they are not simplicial complexes. They are, in fact, the trivalent dual graphs of the octahedron and icosahedron, respectively, while the tetrahedron is self-dual.}. We are therefore unable to construct arbitrarily refined discretizations of a sphere which satisfy the constraints of uniform circumradius and perimeter exactly. Instead, we will generate a sequence of lattices such that the non-uniformity in these two quantities goes to zero in the continuum limit.

We define the non-uniformity in the circumradius $R$ and perimeter $P$ as
\begin{equation}
    E_{R} = \dfrac{\langle R^2 \rangle}{\langle R \rangle^2} - 1 \qquad \text{and} \qquad E_{P} = \dfrac{\langle P^2 \rangle}{\langle P \rangle^2} - 1
\end{equation}
where the angle brackets denote an average over all triangles in the simplicial complex. These quantities can be understood as the normalized variance in the circumradius and perimeter over the entire lattice. Our goal is to move the vertices of the basic discretization (without adding or removing any edges) so that both of these quantities are minimized while retaining the point group symmetry of the original lattice. We do this by minimizing the sum $E = E_R + E_P$. In general, there are not enough degrees of freedom to find a solution such that $E=0$, however we will show that the minimum value of $E$ goes to zero as the square of the effective lattice spacing. As shown in Chapter \ref{chapter:sphere}, this is sufficient to restore rotational symmetry of the critical Ising model in the continuum limit.

\section{Vertex degrees of freedom}

In order to retain as much symmetry as possible, we would like to modify the basic discretization of $S^2$ without breaking the discrete rotational symmetries of the original lattice. This can be accomplished by a judicious choice of the degrees of freedom of the problem, which I will describe here for an icosahedral discretization. It is important to note that the simplicial graph remains fixed during this procedure. This method works equally well for discretizations with octahedral or tetrahedral symmetry, and can also be generalized for discretizations of $S^3$. It is also possible to generalize to non-spherical manifolds, though we have not yet had a reason to do so.

We first identify ``orbits'' which are sets of vertices which transform into one another under the icosahedral group transformations. We use the full icosahedral group which includes rotations and reflections (120 group elements), but this method will also work for the chiral icosahedral group which includes only rotations (60 group elements). The number of distinct vertices in an orbit (the orbit's degeneracy) depends on whether its vertices lie on any symmetry axes of the icosahedron. For example, an orbit with vertices at the midpoints of the icosahedral edges has a degeneracy of 30 (one per icosahedral edge), while an orbit with vertices on one of the reflection axes of an icosahedral face has a degeneracy of 60 (three per icosahedral face). An orbit which does not lie on any symmetry axes has a degeneracy of 120 (six per icosahedral face).

We parameterize an orbit's position by its barycentric coordinates within an icosahedral face. The barycentric coordinates describe a position on an icosahedral face, which is then projected onto the unit sphere. Using barycentric coordinates allows us to define the coordinates of all vertices in an orbit simultaneously on all 20 icosahedral faces. In addition, permuting the order of the 3 barycentric coordinate values generates all of the vertices within a single icosahedral face.

In order to preserve icosahedral symmetry, we require that orbits with vertices on a symmetry axis remains on that symmetry axis during the minimization procedure. Thus, though a point on a sphere has two degrees of freedom in general, this constraint reduces the number of degrees of freedom for some orbits (e.g. an orbit with vertices at the midpoints of the icosahedral edges has no degrees of freedom). The reduced set of orbit degrees of freedom, denoted $\xi_i$ where $i$ runs over all of the remaining degrees of freedom, can be freely adjusted without breaking icosahedral symmetry. Once we determine the values of $\xi_i$ which minimize $E$, we simply use the action of the icosahedral group elements to compute the vertex coordinates of all of the lattice sites in each orbit as described above.

\section{Non-linear solver}

After identifying the reduced set of degrees of freedom $\xi_i$, we use Newton's method to solve the non-linear system of equations
\begin{equation}
    \dfrac{\partial E}{\partial \xi_i} = 0\;.
\end{equation}
We use the barycentric coordinates of the vertices from the basic discretization as the initial guess $\xi_i^{(0)}$, then the $(k+1)$-th iteration of Newton's method~\cite{Dennis1983NumericalMF} sets
\begin{equation}
    \xi_i^{(k+1)} = \xi_i^{(k)} - \left[ \dfrac{\partial^2 E^{(k)}}{\partial \xi_i^{(k)} \partial \xi_j^{(k)}} + \mu^{(k)} \delta_{ij} \right]^{-1} \dfrac{\partial E^{(k)}}{\partial \xi_j^{(k)}}
\end{equation}
where $E^{(k)}$ is the non-uniformity measure computed on the lattice with the vertex coordinates determined by the orbit coordinates $\xi_i^{(k)}$. We compute the partial derivatives via first-order finite differences with a step size of $10^{-5}$, which is efficient for double-precision floating point numbers. The parameter $\mu^{(k)}$ is a preconditioning factor and is chosen to ensure that $E^{(k)}$ is strictly decreasing for successive iterations. We continue the iterative process until the quantity $1 - E^{(k+1)} / E^{(k)}$ becomes less than $10^{-10}$.

\begin{figure}
    \centering
    \includegraphics[width=0.49\textwidth]{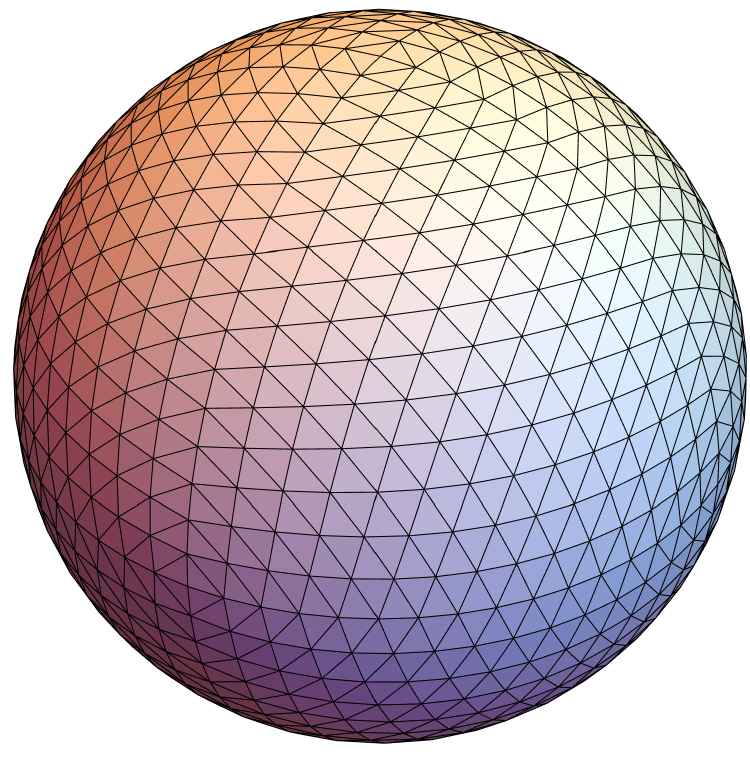}
    \includegraphics[width=0.49\textwidth]{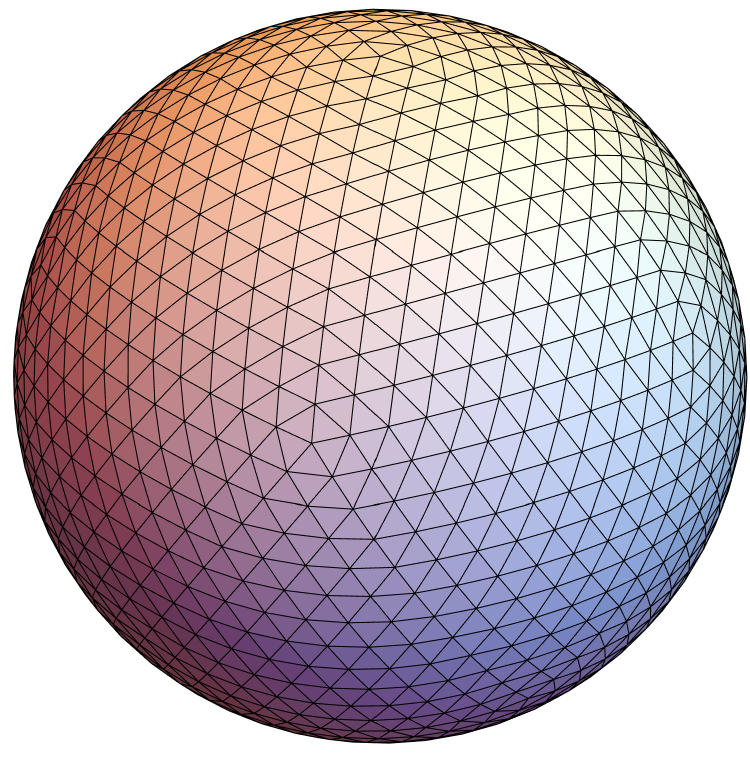}
    \caption{Simplicial discretizations of the sphere after applying the optimization procedure described in this appendix. On the left is the octahedral lattice with a refinement of 16, and on the right is the icosahedral lattice with a refinemenet of 12.}
    \label{fig:lattice_uniform}
\end{figure}

\begin{figure}
    \centering
    \includegraphics{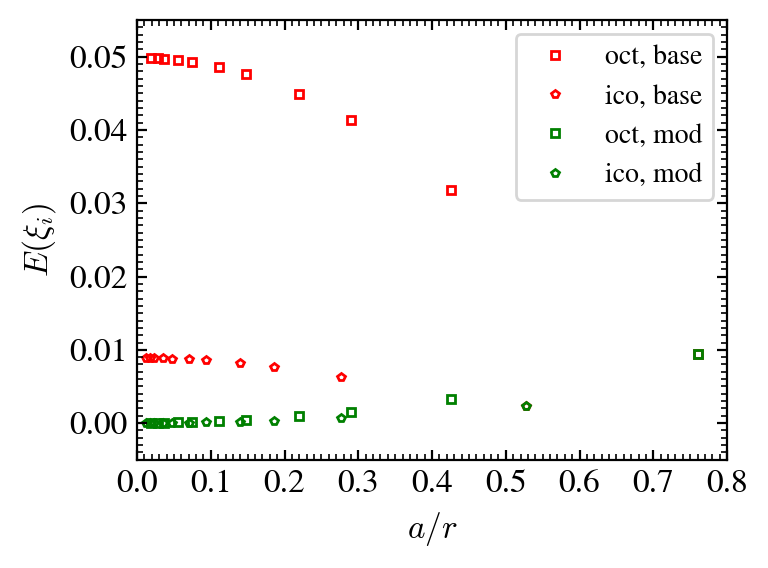}
    \caption{Non-uniformity measure for the octahedral and icosahedral discretizations of $S^2$ as a function of lattice spacing before and after being modified using the optimization procedure described in this appendix.}
    \label{fig:uniform}
\end{figure}

In Fig. \ref{fig:lattice_uniform} we show examples of modified octahedral and icosahedral discretizations of $S^2$. For a quantitative comparison of the unmodified basic lattice and the modified lattice, in Fig. \ref{fig:uniform} we show the non-uniformity measure $E$ as a function of the lattice spacing as defined in Eq. (\ref{eq:a_lat_def}) for both the octahedral and icosahedral lattices. We can clearly see that the non-uniformity in the unmodified lattice approaches a nonzero value in the continuum limit, whereas in the modified lattice the non-uniformity goes to zero roughly quadratically in the lattice spacing.

\end{appendices}
\newpage
\singlespace
\bibliographystyle{apalike}

\bibliography{thesis}
\cleardoublepage

\addcontentsline{toc}{chapter}{Curriculum Vitae}

\begin{center}
{\LARGE {\bf CURRICULUM VITAE}}\\
\vspace{0.5in}
{\large {\bf Evan Owen}}
\end{center}

\noindent
{\bf Education \& Work Experience}
\vspace{1em}

\noindent
Boston University \hfill 1/2021 -- 6/2023 \\ Ph.D. Physics, Advisor: Rich Brower
\vspace{1em}

\noindent
University of Colorado Boulder \hfill 8/2019 -- 12/2020 \\Ph.D. Student
\vspace{1em}

\noindent
San Francisco State University \hfill 1/2017 -- 6/2019 \\ M.S. Physics, Advisor: Jeff Greensite
\vspace{1em}

\noindent
Sir Francis Drake High School (San Anselmo, CA) \hfill 8/2015 -- 6/2016 \\
Math Teacher
\vspace{1em}

\noindent
Chevron Richmond Refinery (Richmond, CA) \hfill 4/2009 -- 8/2019 \\ Design Engineer
\vspace{1em}

\noindent
California Polytechnic State University, San Luis Obispo \hfill 9/2004 -- 3/2009 \\ B.S. Mechanical Engineering
\vspace{1em}

\noindent
{\bf Publications \& Invited Talks}
\vspace{1em}

\noindent
E. Owen, {\em The modular Ising model}, MIT Virtual Lattice Colloquium Series (2022)
\vspace{1em}

\noindent
R.C. Brower and E. Owen, {\em The critical Ising model on an affine plane}, The 39th International Symposium on Lattice Field Theory (Lattice 2022), arXiv 2209:15546
\vspace{1em}

\noindent
R.C. Brower, C.V. Cogburn, and E. Owen, {\em Hyperbolic lattice for scalar field theory in AdS$_3$}, Phys. Rev. D {\bf 105}, 114503 (2022)
\vspace{1em}

\noindent
C. Berger, R.C. Brower, G.T. Fleming, A. Gasbarro, E. Owen and T. Raben, {\em Quantum Counter-Terms for Lattice Field Theory on Curved Manifolds}, PoS(LATTICE2021) {\bf 396}, 215 (2022)
\vspace{1em}

\noindent
J. Greensite and E. Owen, {\em Screening without screening: Baryon energy at high baryon density}, Phys. Rev. D {\bf 100}, 034503 (2019)
\vspace{1em}

\noindent
J. Greensite and E. Owen, {\em Unphysical properties of the static quark-antiquark four-point correlator in Landau gauge}, Phys. Rev. D {\bf 99}, 014506 (2019)

\end{document}